\documentclass{svjour3}
\usepackage[utf8]{inputenc}
\usepackage{graphicx}
\usepackage{natbib}
\usepackage{caption}
\usepackage{amsmath}
\usepackage{aas_macros}
\usepackage{tabularx}
\usepackage{color}
\newcommand{\der}[1] {\mathrm{d}#1}
\newcommand{\derk}[1] {\mathrm{d}^3#1}

\def\makeheadbox{\relax}
\begin{document}
\title{Vlasov methods in space physics and astrophysics 
}


\author{Minna Palmroth         \and
        Urs Ganse              \and
        Yann Pfau-Kempf        \and
        Markus Battarbee       \and 
        Lucile Turc            \and 
        Thiago Brito           \and 
        Maxime Grandin         \and
        Sanni Hoilijoki        \and
        Arto Sandroos	       \and
        Sebastian von Alfthan
}

\authorrunning{M. Palmroth \textit{et al.}} 

\institute{M. Palmroth \at
              Department of Physics, University of Helsinki, Helsinki, Finland; also
              \at Finnish Meteorological Institute, Helsinki, Finland \\
              \email{minna.palmroth@helsinki.fi}           
           \and
           U. Ganse, Y. Pfau-Kempf, M. Battarbee, L. Turc, T. Brito, M. Grandin \at
              Department of Physics, University of Helsinki, Helsinki, Finland 
           \and 
           Sanni Hoilijoki \at
           	  Laboratory for Atmospheric and Space Plasma Physics, University of Colorado at Boulder, Boulder, USA 
           \and 
           Arto Sandroos \at
           Cadence Design Systems, San Jose, CA, USA
           \and
           Sebastian von Alfthan \at
           	  CSC -- IT Center for Science, Espoo, Finland
}

\date{Received: 21.03.2018 / Accepted: 06.07.2018 / Published: 16.08.2018}

\maketitle


\begin{abstract}
This paper reviews Vlasov-based numerical methods used to model plasma in space physics and astrophysics. Plasma consists of collectively behaving charged particles that form the major part of baryonic matter in the Universe. Many concepts ranging from our own planetary environment to the Solar system and beyond can be understood in terms of kinetic plasma physics, represented by the Vlasov equation. We introduce the physical basis for the Vlasov system, and then outline the associated numerical methods that are typically used. A particular application of the Vlasov system is Vlasiator, the world's first global hybrid-Vlasov simulation for the Earth's magnetic domain, the magnetosphere. We introduce the design strategies for Vlasiator and outline its numerical concepts ranging from solvers to coupling schemes. We review Vlasiator's parallelisation methods and introduce the used high-performance computing (HPC) techniques. A short review of verification, validation and physical results is included. The purpose of the paper is to present the Vlasov system and introduce an example implementation, and to illustrate that even with massive computational challenges, an accurate description of physics can be rewarding in itself and significantly advance our understanding. Upcoming supercomputing resources are making similar efforts feasible in other fields as well, making our design options relevant for others facing similar challenges.


\end{abstract}

\section{Introduction}

While physical understanding is inherently based on empirical evidence, numerical simulation tools have become an integral part of the majority of fields within physics. When tested against observations, numerical models can strengthen or invalidate existing theories and quantify the degree to which the theories have to be improved. Simulation results can also complement observations by giving them a larger context. In space physics, spacecraft measurements concern only one point at one time in the vast volume of space, indicating that discerning spatial phenomena from temporal changes is difficult. This is a shortcoming that has also led to the use of spacecraft constellations, like the European Space Agency's Cluster mission \citep{Escoubet2001AnnGeo}. However, simulations are considerably more cost-effective compared to spacecraft, and they can be adopted to address physical systems that cannot be reached by \textit{in situ} experiments, like the distant galaxies. Finally, and most importantly, predictions of physical environments under varying conditions are always based on modelling. Predicting the near-Earth environment in particular has become increasingly important, not only because the near-Earth space hosts expensive assets used to monitor our planet. The space environmental conditions threatening space- or ground-based technology or human life are commonly termed as \textit{space weather}. Space weather predictions include two types of modelling efforts; those targeting real-time modelling (similar to terrestrial weather models), and those which test and improve the current space physical understanding together with top-tier experiments. This paper concerns the latter approach.

The physical conditions within the near-Earth space are mostly determined by physics of collisionless plasmas, where the dominant physical interactions are caused by electromagnetic forces over a collection of charged particles. There are three main approaches to model plasmas: 1) the fluid approach (e.g., magnetohydrodynamics, MHD), 2) the fully kinetic approach, and 3) hybrid approaches combining the first two. Present global models including the entire near-Earth space in three dimensions (3D) and resolving the couplings between different regions are largely based on MHD \citep[e.g.][]{Janhunen2012}. However, single-fluid MHD models are basically scale-less in that they assume that plasmas have a single temperature approximated by a Maxwellian distribution. Therefore they provide a limited context to the newest space missions, which produce high-fidelity multi-point observations of spatially overlapping multi-temperature plasmas. The second approach uses a kinetic formulation as represented by the Vlasov theory \citep{Vlasov}. In this approach, plasmas are treated as velocity distribution functions in a six-dimensional phase space consisting of three-dimensional ordinary space (3D) and a three-dimensional velocity space (3V). \label{sec:pic}The majority of kinetic simulations model the Vlasov theory by a particle-in-cell (PIC) method (Lapenta, 2012\nocite{LAPENTA2012}; Cerutti et al., Living Reviews in Computational Astronomy, in preparation), where a large number of particles are propagated within the simulation, and the distribution function is constructed from particle statistics in space and time. The fully kinetic PIC approach means that both electrons and protons are treated as particles within the simulation. Such simulations in 3D are computationally extremely costly, and can only be carried out in local geometries \citep[e.g.][]{Daughton11}.

A hybrid approach in the kinetic simulation regime means usually that electrons are treated with a fluid description, but protons and heavier ions are treated kinetically. Again, the vast majority of simulations use a hybrid-PIC approach, which have previously concidered 2D spatial regimes due to computational challenges \citep[e.g.,][]{omidi05, karimabadi14}, but have recently been extended into 3D using a limited resolution \citep[e.g.][]{lu15, lin17}. This paper does not discuss the details of the PIC approach, but instead concentrates on a hybrid-Vlasov method, where the ion velocity distribution is discretised and modelled with a 3D-3V grid. The difference to hybrid-PIC is that in hybrid-Vlasov the distribution functions are evolved in time as an entity, and not constructed from particle statistics. The main advantage is therefore that the distribution function becomes noiseless. This can be important for the problem at hand, because the distribution function is in many respects the core of plasma physics as the majority of the plasma parameters and processes can be derived from it. As will be described, hybrid-Vlasov methods have been used mostly in local geometries, because the 3D-3V requirement implies a large computational cost. A global approach, which in space physics means simulation box sizes exceeding thousands of ion inertial lengths or gyroradii per dimension, have not been possible as naturally the large volume has to consider the velocity space as well. The world's (so far) only global magnetospheric hybrid-Vlasov simulation, the massively parallel Vlasiator, is therefore the prime application in this article.

This paper is organised as follows: Section \ref{sec:physicalSystems} introduces the typical plasma systems and relevant processes one encounters in space. Sections \ref{sec:vlasoveq} and \ref{sec:numerical} introduce the Vlasov theory and its numerical representations. Section \ref{sec:Vlasiator} describes Vlasiator in detail and justifies the decisions made in the design of the code to aid those who would like to design their own (hybrid-)Vlasov system. At the time of writing, there are no standard verification cases for a (hybrid-)Vlasov system, but we describe the test cases used for Vlasiator. The physical findings are then illustrated briefly, showing that Vlasiator has made a paradigm change in space physics, emphasising the role of scale coupling in large-scale plasma systems. While this paper concerns mostly the near-Earth environment, we hope it is useful for astrophysical applications as well. Astrophysical large-scale modelling is still mostly based on non-magnetised gas \citep{springel2005,Bryan2014}, while in reality astrophysical objects are in the plasma state. In the future, pending new supercomputer infrastructure, it may be possible to design astrophysical simulations based on MHD first, and later possibly on kinetic theories. If this becomes feasible, we hope that our design strategies, complemented and validated by \textit{in situ} measurements, can be helpful.

\section{Kinetic physics in astrophysical plasmas}
\label{sec:physicalSystems}

Thermal and non-thermal interactions between charged particles and electromagnetic fields follow the same basic rules throughout the universe, but the applicability of simplified theories and the relevant spatial, temporal, and virial scales vary greatly between different scopes of research. In this section, we present an overview of regions of interest and the phenomena found within them.

\subsection{Astrophysical media and objects}
\label{subsec:astrophysical}

Prime examples of themes requiring modelling are e.g., the dynamics of hot, cold, and dark matter in an expanding universe with unknown boundaries. The birth of the universe connects the rapid expansion and cooling of baryonic matter with quantum fluctuation anisotropies that eventually lead to the formation of galactic superclusters. Astrophysical simulations of the universe should naturally account for expansion of space-time and associated effects of general relativity, and modelling of high-energy phenomena should correctly account for special relativity due to velocities approaching the speed of light.
A recent forerunner in modelling the universe is EAGLE \citep{2015MNRAS.446..521S}, which utilises smoothed particle hydrodynamics, with subgrid modelling providing feedback of star formation, radiative cooling, stellar mass loss and feedback from stars and accreting black holes. These simulations operate on very much larger scales compared to the Vlasov equation for ions and electrons, yet they depend strongly on knowledge of processes at smaller length and time scales. Due to the majority of the universe consisting of the mostly empty interstellar and intergalactic media, the energy content of turbulent space plasmas must be understood. This has been investigated through the Vlasov equation \citep[see, e.g.,][]{Weinstock1969}. Conversely, turbulent behaviour at large scales can act as a model for extending power laws to smaller scales \citep{Maier2009}. An alternative if less common approach for modelling galactic dynamics is to describe the distribution of stars as a Vlasov-Poisson system, to be explained below, with gravitational force terms instead of electromagnetic effects \citep{Guo2008}. This approach highlights the use of the Vlasov equation also on large spatial scales.

\subsection{Solar system}

Plasma simulations of the solar system are mostly concerned with the modelling of solar activity and its influence on the heliosphere. Solar activity can be divided into two components: the solar wind, consisting of particles escaping continuously from the solar corona due to its thermal expansion, and carrying with them turbulent fields; and transient phenomena such as flares and coronal mass ejections, during which energy and plasma are released explosively from the Sun into the heliosphere. Topics of active research in solar physics include for example the acceleration and expansion of the solar wind \citep{Yang2012,Verdini2016,Pinto2017}, coronal heating \citep{DeMoortel2014,Cranmer2017} and flux emergence \citep{Schmieder2014}. The latter is particularly important for transient solar activity, as flares and coronal mass ejections are due to the destabilisation of coronal magnetic structures through magnetic reconnection. The typical length of these coronal structures ranges between $10^{6}$ and $10^{8}$ m. Of great interest is also the propagation of the solar wind and solar transients into the heliosphere, in particular for studying their interaction with Earth and other planetary environments. Because of the large scales of the systems considered, solar and heliospheric simulations are generally based on MHD, indicating that currently existing theories of the Sun and the solar eruption are mostly based on the MHD approximation. Applying the Vlasov approach to near-Earth physics, having important analogies to the solar plasmas, may therefore provide important feedback to existing solar theories as well.

\subsection{Near-Earth space and other planetary environments}

\begin{figure}
   \centering
   \includegraphics[width=\textwidth]{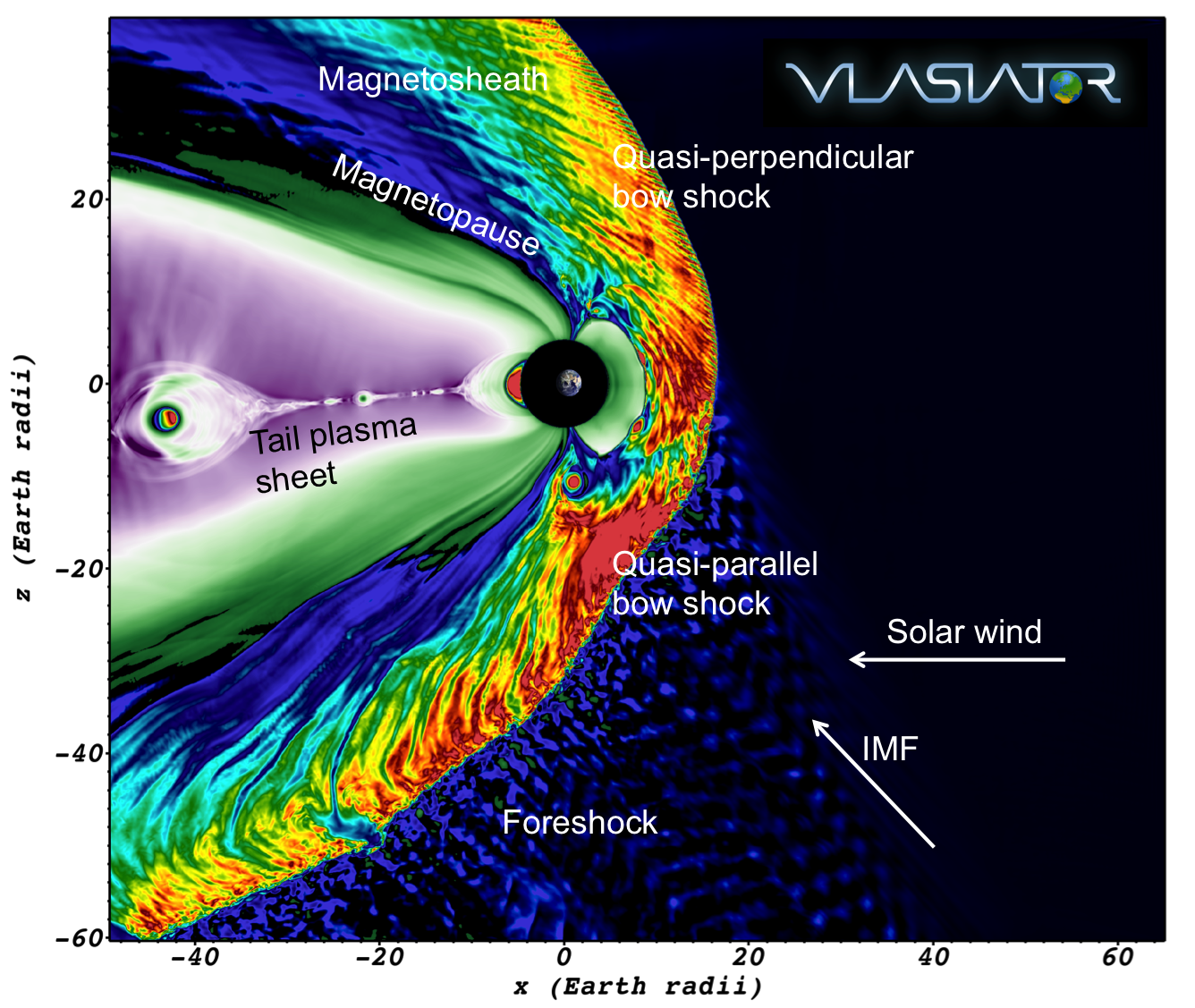}
   \caption{Vlasiator modelling of the magnetosphere in the noon-midnight meridian plane viewed from the morning sector. Different physical regions outlined in the text are annotated, along with the solar wind and IMF directions.}
   \label{fig:msphere}
\end{figure}

Figure \ref{fig:msphere} illustrates the near-Earth space. The shock separating the terrestrial magnetic domain from the solar wind is called the \textit{bow shock} \citep[e.g.][]{omidi95}, and the region of shocked plasma downstream is the \textit{magnetosheath} \citep{Balogh13}. The interplanetary magnetic field (IMF), which at 1\,AU typically forms an angle of 45$^{\circ}$ relative to the plasma flow direction, intensifies at the shock, increasing the magnetic field strength to roughly four-fold compared to that in the solar wind \citep[e.g.][]{spreiter94}.

The bow shock--magnetosheath system hosts highly variable and turbulent environmental conditions, with the bow shock normal angle with respect to the IMF direction being one of the most important factors controlling the level of variability. At portions of the bow shock where the IMF is quasi-parallel with the bow shock normal (termed \textit{quasi-parallel shock}), some particles reflect at the shock and propagate back upstream causing instabilities and waves in the \textit{foreshock} upstream of the bow shock \citep[e.g.][]{hoppe81}. On the quasi-perpendicular side of the shock, where the IMF direction is more perpendicular to the bow shock normal, the downstream magnetosheath is much smoother, but exhibits large-scale waves originating from anisotropies in the ion distribution function \citep[e.g.][]{Genot2011, soucek15, Hoilijoki2016}. The foreshock--bow shock--magnetosheath coupled system is under active research, and since it is the magnetosheath plasma which ultimately determines the conditions within the near-Earth space, most important open questions include the processes which determine the plasma characteristics in space and time. The entire system has previously been modelled with MHD, which is usable to infer average properties of the dayside system \citep[e.g.][]{palmroth01, Chapman2003,Dimmock2013,Mejnertsen2018}, but unable to take into account particle reflection, kinetic waves, turbulence, and it neglects e.g., plasma asymmetries between the quasi-parallel and quasi-perpendicular sides of the shock that require a non-Maxwellian ion distribution function.

The earthward boundary of the magnetosheath is called \textit{the magnetopause}, a current layer exhibiting large gradients in the plasma parameter space. Energy and mass exchange between the upstream plasma and the magnetosphere occurs at the magnetopause \citep{palmroth03, palmroth06, pulkkinen06, anekallu11,Daughton2014,Nakamura2017}, and therefore its processes are important in determining the amount of energy driving the space weather phenomena, which can endanger technological systems or human health \citep{Watermann2009, Eastwood2017}. Space weather phenomena are complicated and varied, and we give a non-exhaustive list just to name a few most important categories. Direct energetic particle flows from the Sun alter the communication conditions especially at high latitudes, affecting radio broadcasts, aircraft communication with air traffic control, and radar signals. Sudden changes in the magnetic field induce currents in the terrestrial long conductors, such as gas pipelines, railways, and power grids that can sometimes be disrupted \citep[e.g.][]{wik08}. Increasing numbers of satellites are being launched, vulnerable to sudden events in the geospace, as it has been experienced that some spacecraft have stopped operation in response to space weather events \citep{Green2017}. Overall, some estimations show that in the worst case, an extreme space weather event could induce economic costs of the order of 1 to 2~trillion USD during the first year following its occurrence, and that it could take 4 to 10~years for the society to recover from its effects \citep{NAP12507}. Understanding and predicting the geospace is ultimately done by modelling. While the previous global MHD models can be executed near real-time and they provide the average description of the system, they cannot capture the kinetic physics that is needed to explain the most severe space weather events.

One additional factor in the accurate modelling of the geospace as a global system is that one needs to address the ionised upper atmosphere called \textit{the ionosphere} within the simulation. The Earth's ionosphere is a weakly ionised medium, divided into three regions -- named $D$ (60--90~km), $E$ (90--150~km), and $F$ ($>$150~km) -- corresponding to three peaks in the electron density profile \citep{Hargreaves_textbook}. From the magnetospheric point of view, the ionosphere represents a conducting layer closing currents flowing between the ionosphere and magnetosphere \citep{Merkin2010}, reflecting waves \citep{Wright2014}, and depositing precipitating particles \citep[e.g.,][]{Rodger2013}. Further, the ionosphere is a source of cold electrons \citep{CranMcGreehin2005} and heavier ions \citep[e.g.,][]{peterson81}. These cold ions of ionospheric origin may affect local processes in the magnetosphere, such as magnetic reconnection at the magnetopause \citep{Andre2010,ToledoRedondo2016}. The global MHD models typically use an electrostatic module for the ionosphere, coupled to the magnetosphere by currents, precipitation and electric potential \citep[e.g.,][]{Janhunen2012, palmroth06prec}. The ionosphere itself is modelled either empirically or based on first principles: The International Reference Ionosphere (IRI) model describes the ionosphere empirically from 50~km to 1500~km altitude \citep{Bilitza2008}, while for instance, the Sodankyl\"a Ion and Neutral Chemistry model solves the photochemistry of the $D$ region, taking into account several hundred chemical reactions involving 63~ions and 13 neutral species \citep[][and references therein]{Verronen2005}. At higher altitudes, transport processes become important, and models such as TRANSCAR \citep[][and references therein]{Blelly2005} or the IRAP Plasmasphere--Ionosphere Model \citep{Marchaudon2015} couple a kinetic model for the transport of suprathermal electrons with a fluid approach to resolve the chemistry and transport of ions and thermal electrons in the convecting ionosphere. Neither empirical nor the first-principles based models are using the Vlasov equation, which at the ionosphere concerns much finer scales. 

In general, the interaction of the solar wind with the other magnetized planets in our solar system is essentially similar to that with Earth. The main differences stem from the scales of the systems, which depend on the strength of their intrinsic magnetic field and the solar wind parameters changing with heliospheric distance. While the modelling of the magnetospheres of the outer giants is only achievable to date using fluid approaches, the small size of Mercury's magnetosphere has been targeted for global kinetic simulations \citep{Richer2012}. For the same reason, kinetic models are also a popular tool to investigate the plasma environment of non-magnetized bodies such as Mars, Venus, comets, and asteroids. In particular, \citet{Umeda2014} and \citet{Umeda2015} have studied the interaction of a weakly magnetized body with the solar wind by means of full-Vlasov simulations.

\subsection{Scales and processes}

The following processes are central in explaining plasma behaviour in the Solar-Terrestrial system and astrophysical domains: 1) magnetic reconnection enabling energy and mass transfer between different magnetic domains, 2) shocks forming due to supersonic relative flow speeds between plasma populations, 3) turbulence providing energy dissipation across scales, and 4) plasma instabilities transferring energy between the plasma and waves. All these processes contribute to particle acceleration, which is one of the most researched topics within Solar-Terrestrial and astrophysical domains, and notorious in requiring understanding of both local microphysics and global scales. Below, we introduce some examples of these processes within systems having scales that can be addressed with the Vlasov approach. Simulations of non-thermal space plasmas encompass a vast range of scales, from the smallest ones (electron scales, ion kinetic scales) to local and even global structures. Table \ref{tab:Scales} lists typical ranges of a handful of plasma parameters encountered in different branches of space sciences and astrophysics. Especially in a larger astrophysical context, simulations cannot directly encompass all relevant spatial and temporal scales. It is important to note, however, that scientific results of kinetic effects can be achieved even without directly resolving all the spatial scales that may at first glance appear to be a requirement \citep[]{PfauKempf2018}.

Reconnection is a process whereby oppositely oriented magnetic fields break and re-join, allowing a change in magnetic topology, plasma mixing, and energy transfer between different magnetic domains. Within the magnetosphere, reconnection occurs between the terrestrial northward oriented magnetic field and the magnetosheath magnetic field that mostly mimics the direction of the IMF, but  can sometimes be significantly altered due to magnetosheath processes \citep{Turc2017}. Magnetospheric energy transfer is most efficient when the magnetosheath magnetic field is southward, while for northward IMF reconnection locations move to the nightside lobes \citep{palmroth06}. Actively researched topics focus on understanding the nature and location of reconnection as a function of driving conditions \citep[e.g.][]{hoilijoki14, fuselier17}. Energy transfer at the magnetopause sets the near-Earth space into a global circulation \citep{Dungey1961}, leading to reconnection in the magnetospheric \textit{tail}. The tail centre hosts a hot and dense \textit{plasma sheet}, the home of perhaps most diligent scientific investigations within the domain of magnetospheric physics. Especially in focus have been explosive times when the magnetospheric tail disrupts and launches into space, accelerating particles and causing abrupt changes in the global geospace \citep[e.g.][]{Sergeev2012}. Reconnection has been suggested as one of the main drivers of tail disruptions \citep[e.g.][]{angelopoulos08}, while other theories related to plasma kinetic instabilities exist as well \citep[e.g.][]{lui96}. Tail disruptions have important analogues in solar eruptions \citep[e.g.][]{birn09}, and investigating the tail disruptions with global simulations together with \textit{in situ} measurements may shed light into other astrophysical systems as well.

Collisionless shocks form due to plasma populations flowing supersonically with respect to each other, redistributing flow energy into thermal energy and accelerating particles \citep[e.g.][]{Balogh13, Marcowith2016}. Shock fronts such as those found at supernova explosions are an efficient accelerator \citep{1949PhRv...75.1169F}. Diffusive shock acceleration \citep[e.g.,][]{Axford1977,Krymskii1977,Blandford1978,Bell1978} is the primary source of solar energetic particles, and occurs from the non-relativistic \citep[e.g.][]{Lee2005} to the hyper-relativistic \citep{Aguilar2015} energy regimes. Shock-particle interactions including kinetic effects have been modelled using various analytical and semiempirical methods (see, e.g. \citealt{2015A&A...584A..81A,Afanasiev2018,2017JGRA..12210938H,2016ApJ...831..120K,2017JPhCS.900a2013L,2010AdSpR..46....1L,2008ApJ...686L.123N,2009ApJ...696..261S,2014JSWSC...4A..08V}), but drastic approximations are usually required in order to model the whole acceleration process, and Vlasov methods have not yet been utilised. The classic extension of hydrodynamic shocks into the MHD regime has been disproven by a number of hybrid models due to, e.g., shock reformation \citep{Caprioli2013,Hao2017} and anisotropic pressure and energy imbalances due to non-thermal particle populations \citep{Chao1995,Genot2009}. Only a self-consistent treatment including kinetic effects is capable of describing diffusive shock acceleration accurately. Recent works coupling shocks and high-energy particle effects include, e.g., those by \citet{Guo2013,Bykov2014,Bai2015,VanMarle2018}. Challenges associated with simulating shocks include modelling gyrokinetic scales for ions whilst allowing the simulation to cover the large spatial volume involved in the particle trapping and energisation process. Radially expanding shock fronts within strong magnetic domains result in a requirement for high resolution both spatially and temporally. Modern numerical approaches usually make some sacrifices, e.g. performing 1D-2V self-consistent calculations or advancing 3D-1V semi-analytical models. The Vlasov approach is especially interesting in probing the physics of particle injection, trapping, acceleration and escape.

In addition to shock acceleration, kinetic simulations of solar system plasmas are also applied to the study of solar wind turbulence. How energy cascades from large to small scales and is eventually dissipated is an outstanding question, which can be addressed using kinetic simulations \citep{Bruno2013}. Hybrid-Vlasov simulations \citep{Valentini2010,Verscharen2012,Perrone2012} have in particular been utilised to study the fluctuations around the ion inertial scale, which is of particular importance as it marks the transition between the fluid and kinetic scales.

Plasma instabilities arise when a source of free energy in the plasma allows a wave mode to grow non-linearly. They are ubiquitous in our universe, and play an important role in both solar-terrestrial physics, where, for example, the Kelvin-Helmholtz instability transfers solar wind plasma into the Earth magnetosphere \citep[e.g.,][]{Nakamura2017}, and in astrophysical media, for instance in accretion disks, where the turbulence is driven by the 
magnetorotational instability. Vlasov models have been applied to the study of many instabilities, such as the Rayleigh-Taylor instability \citet{Umeda2016RT,Umeda2017RT}, Weibel-type instabilities \citet{Inglebert2011,Ghizzo2017}, and the Kelvin-Helmholtz instability \citet{Umeda2010}.

\begin{table}
\captionsetup{singlelinecheck=off}
   \begin{tabular}{l|c|c|c}
   	  \hline
       Physical system           & Near-Earth space & Solar system & Astrophysics \\
	  \hline
      \hline
      $\lambda_\mathrm{D}$ (m)   & $10^{-3} - 10^2$ & $10^{-4} - 10^{1}$ & $ 10^{-4} - 10^{5}$ \\
      $c/\omega_\mathrm{pi}$ (m) & $<10^5$          & $10^{3} - 10^{5}$  & $10^{-3} - 10^{6}$\\
      $r_\mathrm{Li}$ (m)        & $10^3 - 10^6$    & $10^{1} - 10^{7}$ & $10^{-3} - 10^{8}$ \\
      System size (m)            & $10^9$           & $10^{11}$ &  $10^{15} - 10^{25}$ \\
      Process time scales        & 1\,s -- 1\,d     & 1\,s -- 1\,mo & 1\,s -- 10 Gy \\
      \hline
      \hline
   \end{tabular}
\caption[foo]{Typical plasma parameters and scales for solar-terrestrial and astrophysical phenomena. $\lambda_\mathrm{D}$: the Debye length is the characteristic of a plasma related to its ability to shield out the electric potentials applied to it; $c/\omega_\mathrm{pi}$: the ion inertial length is the scale at which ions decouple from electrons; $r_\mathrm{Li}$: the ion Larmor radius is the radius at which the ion gyrates around the magnetic field. 
} 
\label{tab:Scales}
\end{table}

\section{Modelling with the Vlasov equation}
\label{sec:vlasoveq}

Simulating plasma provides numerous challenges from the modelling perspective. Constructing a perfect representation of the plasma, where every single charge carrier is factored in the equations, would require an immense amount of computational power. Spatial scales required to fully describe the plasma environment range from the microscale Debye length, up to the macroscale of the phenomena one is trying to simulate. The particle density needed is such that performing a fully kinetic simulation of even a low-density plasma self-consistently, using Maxwell's equations for the electric and magnetic fields and Lorentz' equation for the protons and electrons, is out of reach even to present day large supercomputers. Currently, only plasma phenomena that occur in relatively short spatial and temporal scales, such as magnetic reconnection and high frequency waves, are modelled using this approach. For this reason, adopting a continuous resolution of the velocity space and using distribution functions as the main object to be simulated provides a meritorious form of simulating plasmas. 

Plasmas can also be treated as a fluid and the standard way of doing that is using the magnetohydrodynamic (MHD) approximation. This modelling approach is more suitable for large domain sizes where detailed information is not necessary. However, MHD does not offer information about small spatio-temporal scales. Statistical mechanics provides some information about a neutral gas based on assumptions done at the atomic scale. The \emph{kinetic plasma approach} treated in the following is based on this same principle. It describes the plasmas using \emph{distribution functions in phase space} and uses Maxwell's equations and the \emph{Vlasov equation} to advance the fields and the distribution functions, respectively.

\subsection{The Vlasov equation}

In plasmas, as in neutral gases, the dynamical state of every constituent particle can be described by its position ($\mathbf{x}$) and momentum ($\mathbf{p}$) (or velocity $\mathbf{v}$) at a given time $t$. It is also common to separate the different species $s$ in a plasma (electrons, protons, helium ions, etc). Accordingly, the dynamical state of a system of particles of species $s$, at a given time, can be described by a \emph{distribution function} $f_s(\mathbf{x},\mathbf{v},t)$ in 6-dimensional space, also called \emph{phase space}.

The distribution function $f_s(\mathbf{x},\mathbf{v},t)$ represents the phase-space density of the species inside a phase-space volume element of size $\mathrm{d}^3\mathbf{x} \mathrm{d}^3\mathbf{v}$ during a time $\der{t}$ at the ($\mathbf{x},\mathbf{v},t$) point. Hence, in a system with $N$ particles, integrating over the spatial volume $\mathcal{V}_r$  and the velocity volume $\mathcal{V}_v$ (i.e. the entire phase-space volume $\mathcal{V}$) one obtains 
\begin{equation}
  N = \int_{\mathcal{V}_r}\int_{\mathcal{V}_v} f_s(\mathbf{x},\mathbf{v},t) \, \derk{\mathbf{x}} \, \derk{\mathbf{v}}.
  \label{eq:phaseSpace}
\end{equation}

It is important to represent and describe the time evolution of the distribution functions given some external conditions. The Boltzmann equation, 
\begin{equation}
  \frac{\partial f_s}{\partial t} + \vec{v} \cdot \frac{\partial f_s}{\partial \vec{x}} + \frac{\vec{F}}{m} \cdot \frac{\partial f_s}{\partial \vec{v}}  = \left( \frac{\partial f_s}{\partial t} \right)_\mathrm{coll}
      \label{eq:boltzmann}
\end{equation}
uses the distribution function $f_s$ to describe the collective behaviour of a system of particles subject to collisions and external forces $\vec{F}$, where the term on the right-hand side represents the forces acting on particles in collisions. Its derivation starts from the standard equation of motion and also takes Liouville's theorem into account (see \ref{sec:liouville}) and it is therefore valid for any Hamiltonian system. In plasmas, the Lorentz force takes the role of the external force and collisions between particles are often neglected. Taking these two assumptions into consideration, one obtains the \emph{Vlasov equation} \citep{Vlasov}, often called ``the collisionless Boltzmann equation'':
\begin{equation}
  \frac{\partial f_s}{\partial t} + \vec{v} \cdot \frac{\partial f_s}{\partial \vec{x}} + \frac{q_s}{m_s}
  \left( \vec{E} + \vec{v} \times \vec{B} \right) \cdot
  \frac{\partial f_s}{\partial \vec{v}}
      = 0.
      \label{eq:vlasov}
\end{equation}

If a significant part of the plasma acquires high enough kinetic energy, then relativistic effects start to become important. It can be shown that the Vlasov equation \eqref{eq:vlasov} is Lorentz-invariant and therefore holds in such cases if $v$ is simply considered to be the proper velocity~\citep{Thomas16PRE}. There are only very few numerical applications related to space or astrophysics that directly solve the relativistic Vlasov equation (as opposed to the particle-in-cell approach, which solves the same physical system through statistical sampling of particles and their propagation, compare section \ref{sec:pic}) but it is more common in other contexts, such as laser-plasma interaction applications \citep{Martins10,Inglebert2011}. By using a frame that is propagating at relativistic speeds, a Lorentz-boosted frame, the smallest time or space scales to be resolved become larger and the plasma length shrinks due to Lorentz contraction, indicating that the simulation execution times are accelerated.


\subsection{Closing the Vlasov equation system}
In any simulation, it is necessary to couple the Vlasov equation with the field equations to form a closed set of equations. The Vlasov equation deals with the time evolution of the distribution function and uses the electromagnetic fields as input. Thus the fields need to be evolved based on the updated distribution function. There are two main ways of closing the equation set: the electrostatic approach, which uses the Poisson equation to close the system and the electromagnetic approach, which uses the Maxwell equations to that end. They are typically referred to as the Vlasov-Poisson system and the Vlasov-Maxwell system of equations. With appropriate approximations, the system can also be closed without solving the Vlasov equation for all species.

\subsubsection*{The Vlasov-Poisson equations}
The Vlasov-Poisson equations model plasma in the electrostatic limit without a magnetic field (corresponding to the assumption that $v/c \rightarrow 0$ for any relevant velocity $v$ in the system). Thus equation \eqref{eq:vlasov} takes the form
\begin{equation}
  \label{eq:electrostaticVlasov}
  \frac{\partial f_s}{\partial t} + \vec{v} \cdot \frac{\partial f_s}{\partial \vec{x}} + \frac{q_s}{m_s}
  \vec{E} \cdot \frac{\partial f_s}{\partial \vec{v}}
      = 0
\end{equation}
for all species and the system is closed by the Poisson equation
\begin{equation}
  \label{eq:Poisson}
  \nabla^2\Phi + \frac{\rho_q}{\epsilon_0} = 0,
\end{equation}
where $\Phi$ is the electric potential and $\epsilon_0$ is the vacuum permittivity. Using equation~\eqref{eq:phaseSpace}, the total charge density $\rho_q$ is obtained by taking the zeroth moment of $f$ for all species:
\begin{equation}
  \rho_q = \frac{1}{\mathcal{V}_r} \sum_s q_s N_s = \sum_s q_s \int_{\mathcal{V}_v} f_s(\mathbf{x},\mathbf{v},t) \, \derk{\mathbf{v}}.
\end{equation}

\subsubsection*{The Vlasov-Maxwell equations}
In the electromagnetic case, the Vlasov equation \eqref{eq:vlasov} is retained for all species and complemented by the Maxwell equations, namely the Amp\`ere law
\begin{equation}
  \nabla\times\mathbf{B} = \mu_0\mathbf{j} + \frac{1}{c^2}\frac{\partial \mathbf{E}}{\partial t},
\label{eq:Ampere}
\end{equation}
the Faraday law
\begin{equation}
  \nabla\times\mathbf{E} = -\frac{\partial \mathbf{B}}{\partial t},
\label{eq:Faraday}
\end{equation}
and the Gauss laws
\begin{equation}
  \nabla\cdot\mathbf{B} = 0 \mathrm{~and~} 
  \nabla\cdot\mathbf{E} = \frac{\rho_q}{\epsilon_0}.
\label{eq:Gauss}
\end{equation}
Usually in a numerical scheme only equations \eqref{eq:Ampere}$-$\eqref{eq:Faraday} are discretised. If equation~\eqref{eq:Gauss} is not satisfied by the numerical method used, numerical instabilities can occur because the underlying system needs to be divergence-free.

\subsubsection*{Hybrid-Vlasov systems}
\label{txt:hybrid_vlasov}
The hybrid-Vlasov systems retain only the Vlasov equation for the ions, thus neglecting the electrons to a certain extent. This has the advantage that the model is not required to resolve the short temporal and spatial scales associated with electron dynamics. Typically, the system is  closed by taking moments of the Vlasov equation and making approximations pertinent to the simulation system at hand.

Integrating \eqref{eq:vlasov} over the velocity space, one gets the \emph{continuity equation} or the zeroth moment of the Vlasov equation:
\begin{equation}
  \frac{\partial n_s}{\partial t} + \nabla \cdot \left( n_s \vec{V}_s \right)
      = 0,
      \label{eq:continuity}
\end{equation}
where $n_s$ and $\vec{V}_s$ are the number density and the bulk velocity of species $s$, respectively. Multiplying \eqref{eq:vlasov} by the phase-space momentum $m_s \vec{v}_s$ and integrating over the velocity space, one obtains the equation of motion or the first moment of the Vlasov equation:
\begin{equation}
  n_s m_s \left( \frac{\partial \vec{V}_s}{\partial t} + (\vec{V}_s \cdot \nabla) \vec{V}_s \right) - n_s q_s
  \left( \vec{E} + \vec{V}_s \times \vec{B} \right) + \nabla \cdot \mathcal{P}_s =
  0,
\label{eq:motion}
\end{equation}
where $\mathcal{P}_s$ is the pressure tensor of species $s$, which can in turn be obtained as the second moment of $f_s$. This leads to a chain of equations where each step depends on the next moment of $f_s$. The most common closure of hybrid-Vlasov systems is taken at this level by summing the electron and ion equations of motion and neglecting terms based on the electron-to-ion mass ratio $m_\mathrm{e}/m_\mathrm{i} \ll 1$, leading to the generalised Ohm's law
\begin{equation}
  \mathbf{E} + \mathbf{V}\times\mathbf{B} = \frac{\mathbf{j}}{\sigma} + \frac{\mathbf{j}\times\mathbf{B}}{n_e e} - \frac{\nabla\cdot\mathcal{P}_\mathrm{e}}{n_e e} + \frac{m_\mathrm{e}}{n_e e^2}\frac{\partial \mathbf{j}}{\partial t},
\label{eq:Ohm}
\end{equation}
where $\sigma$ is the conductivity, $e$ is the elementary charge, $n_e$ is the electron number density, and $\vec{j}$ is the total current density. In the limit of slow temporal variations, the rightmost electron inertia term is typically dropped from the equation. Further, assuming high conductivity, one is left with the Hall term $\mathbf{j}\times\mathbf{B}/\rho_q$ and the electron pressure gradient term $\nabla\cdot\mathcal{P}_\mathrm{e}/\rho_q$ on the right-hand side. The electron pressure term can be handled in a number of ways, such as using isobaric or isothermal assumptions. If electrons are assumed to be completely cold, the equation can be written as the Hall MHD Ohm's law
\begin{equation}
\mathbf{E} + \mathbf{V}\times\mathbf{B} = \frac{\mathbf{j}\times\mathbf{B}}{n_e e}.
\label{eq:HallMHD}
\end{equation}

Thus the hybrid-Vlasov system of equations retains the Vlasov equation \eqref{eq:vlasov} for ions and the Maxwell-Amp\`ere and Maxwell-Faraday equations \eqref{eq:Ampere} and \eqref{eq:Faraday} but replaces the Gauss equations by a generalised Ohm equation \eqref{eq:Ohm} with appropriate approximations. If rapid field fluctuations are excluded from the solution, the displacement current from Amp\`ere's law can be omitted, resulting in the \emph{Darwin approximation} and yielding
\begin{equation}
  \nabla\times\mathbf{B} = \mu_0\mathbf{j}.
\label{eq:AmpereDarwin}
\end{equation}
This makes the equation system more easily tractable.

Note that conversely, neglecting ion dynamics and retaining the electron Vlasov equation can be advantageous in certain contexts and is formally equivalent to the above with switched electron and ion variables.

\subsection{Properties of the Vlasov equation}

When solving the Vlasov equation, there are a number of useful properties that can be used for an advantage in numerical solvers. In its fundamental structure, the Vlasov equation is a 6D advection equation, which equals a zero material derivative of phase-space density. In the absence of any source terms, finding a solution at any given point in time requires to determine the motion of the phase-space density. One particularly handy property follows from \label{sec:liouville} Liouville's theorem, from which the Vlasov equation is derived. It states that \emph{phase-space density is constant along the trajectories of the system}. This means that a solution of the system at one point in time can be followed forward or backward in time arbitrarily, as long as the trajectories of phase-space elements, characterising the Vlasov equation, are known. 

\label{sec:filamentation} One consequence of Liouville's theorem is that initial density perturbations tend to form smaller and smaller structures as trajectories of the phase-space regions with different densities converge over time. In physical reality, this so-called filamentation has a natural cutoff at scales where diffusive scattering effects become important, however, this is not part of the pure mathematical description of the Vlasov equation. Therefore, numerical implementations need to address this issue either by explicit filtering steps, or innate numerical diffusivity.

A fundamental consideration in any physical modelling is the conservation of certain quantities, like mass, momentum, energy, and electric charge. This of course applies to Vlasov-based plasma modelling as well. Variational approaches have been used to derive the Vlasov-Maxwell and Vlasov-Poisson systems of equations as well as reduced forms \citep[e.g.][]{MARSDEN1982,Ye1992,Brizard2000,BrizardTronko2011}. Care has to be taken when developing numerical solutions of the Vlasov equation that quantities relevant to the problem to be solved are conserved adequately by the method \citep[e.g.][]{Filbet2001,Crouseilles2010,Cheng2013,CHENG2014,BecerraSagredo2016,Einkemmer2018}.

\section{Numerical modelling and HPC aspects}
\label{sec:numerical}

Finding solutions to the Vlasov equation in order to model
physical systems typically involves computer simulations. Therefore, the phase space density $f(\vec{x},\vec{v},t)$ needs to be numerically represented, which strongly influences
the choice of solvers and the resulting simulation code. This
section is structured around different numerical representations of $f$.

\label{sec:dimensionality}
A problem common to all numerical approaches solving the Vlasov equation is the
curse of dimensionality -- to fully reproduce all physical behaviour, the simulation
domain must be 6-dimensional, with all 6 dimensions being of sufficient resolution or
fidelity for the desired physical system. This considerably impacts the size of the
phase space and hence the computational burden of the algorithm.

\begin{figure}
	\includegraphics[width=\textwidth]{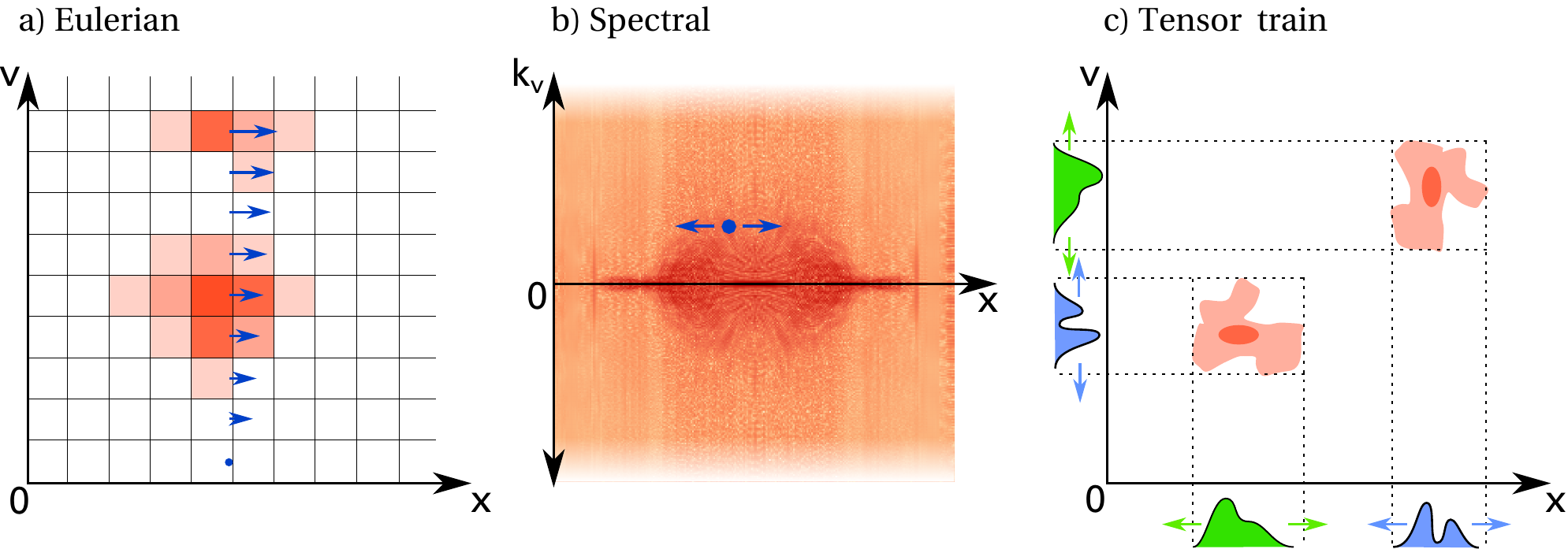}
    \caption{Different ways of numerically representing the phase-space density
    $f(\vec{x},\vec{v},t)$: \textbf{a)} In a Eulerian grid, every grid cell
    stores the local value of phase-space density, which is transported across
    cell boundaries. \textbf{b)} Spectral representations (shown here: Fourier-space 
    in $\vec{v}$) allow for some update steps of phase-space density to be
    performed locally. \textbf{c)} In a tensor train representation, phase-space 
    density is represented as a sum of tensor products of single coordinates'
    distribution functions which get transported individually.
    }
    \label{fig:representations}
\end{figure}

\subsection{Eulerian approach}
\label{sec:eulerian}
In a straightforward manner, the phase-space distribution function $f(\vec{x},\vec{v},t)$ can be 
discretised on a Eulerian grid, which can be operated by different kinds of solvers (see Figure \ref{fig:representations}a). The structure of the Vlasov equation allows both finite volume (FV) as well as semi-Lagrangian solvers to be employed, and all of these have been operated with some success \citep{Arber2002}. Discretisation of velocity space to a finite grid size $\Delta v$ also automatically imposes a lower limit for phase-space filamentation (compare section \ref{sec:filamentation}), at which the grid will naturally smooth out the structure. In some cases this is a purely numerically diffusive process, whereas others use explicit smoothing, filtering or subgrid modelling \citep[e.g.][]{Klimas1987,KlimasFarrell1994}. The 6-dimensional structure of the phase space, along with the physical scales and resolutions imposed by the underlying physical system (compare section \ref{sec:physicalSystems}) make a Eulerian representation in a Cartesian grid computationally impractical in the vast majority of cases concerning a large simulation volume.

Let us consider as an example a simulation of the Earth's entire magnetosphere using a full 3D-3V, Eulerian hybrid-Vlasov model. Let it extend up to the lunar orbit ($x \sim \pm60 \, R_E$ in every direction) resolving approximately the solar wind ion inertial length ($\Delta x \sim 100$ km), and let the velocity space encompass typical solar wind velocities with some safety margin ($v \sim \pm 2000$ km/s) while resolving the solar wind thermal speed ($\Delta v \sim 30$ km/s). In this case the resulting phase space would contain a total of $10^{18}$ cells. If each of them were represented by a single-precision floating point value, a minimum of 4\,EiB of memory would be required to represent it!

Fortunately, there are many possibilities for simplification of the computational grid size:
\begin{itemize}
\item Reduction of phase-space dimension, if the physical system under consideration allows it, is an easy and efficient way to reduce computational load. Simulations of longitudinal wave instabilities \citep{Jenab2014,shoucri2008eulerian} and fundamental studies of filamentation have been performed in a 1D-1V setup, whereas laser wakefield and wave interaction simulations tend to be modelled in 2D-2V or 2D-3V setups \citep{besse4DVlasov, sircombe2004, Thomas16PRE}. Another possibility here is to globally reduce the number of grid points by introducing a sparse grid representation, where the grid may be uneven with respect to Cartesian coordinates, while remaining static during runtime \citep{KormanSparseGrids,Guo2016}. This is sometimes referred to as a sparse grid representation.
    
\item Gyrokinetic simulations reduce the velocity space by dropping the azimuthal velocity dimensions perpendicular to the magnetic field, thus assuming complete gyrotropy of the distribution functions \citep[e.g.][]{Goerler2011JCP}.

\item Adaptively refined grids can be employed to reduce resolution and thus computational expense in areas of phase space that are considered less important for the physical problem at hand \citep{Wettervik2017,Besse2008}.

\item In many physical scenarios, large parts of phase space contain an extremely low, if not zero, density, and contribute nothing to the overall dynamic development. Suitable pruning of the phase-space grid can thus be performed to obtain a data structure that dynamically removes grid elements during runtime and keeps them only in regions deemed relevant for the physical system dynamics. The computational speed-up gained through the reduction of phase-space volume thus obtained can in some cases be a tradeoff against physical accuracy, and needs to be carefully considered. We have implemented this option in Vlasiator, and call it the dynamic  sparse phase space representation, discussed more in section \ref{sec:vlasiatorSparse}. This method is not to be mixed to the static sparse grid methods \citep{KormanSparseGrids,Guo2016} that are fundamentally dimension reduction techniques, similar to the low-rank approximations.

\end{itemize}

In plasmas, the magnetic field $\vec{B}$  makes the particles gyrate while the electric field $\vec{E}$ causes them to accelerate and drift. It can be advantageous to take the characteristics of acceleration due to Lorentz' force into consideration when choosing an appropriate grid for the numerical phase-space representation. Common ideas include:
\begin{itemize}
\item A polar velocity coordinate system aligned with the magnetic field and centred around the drift velocity,
	$\vec{v} = ( v_\parallel,\, v_r,\, v_\phi ),$
	in which the gyrophase coordinate $v_\phi$ has a much lower resolution than $v_\parallel$ and $v_r$. This can be employed in cases where the velocity distribution functions are known not to deviate strongly from gyrotropy, i.e., to exhibit cylindrical symmetry with respect to the magnetic field direction. However, the disadvantage of a polar velocity space over a Cartesian one is the more complex coordinate transformation required for transport into the spatial neighbours.

\item A Cartesian representation of velocity space, in which its coordinate axes co-rotate with the local gyration at every given spatial cell. Such a setup has the advantage that no transformation of velocity space due to the magnetic field will have to be performed  at all, and no numerical diffusion due to gyration motion will occur. It does however come at the cost of more complicated spatial updates, since neighbouring spatial domains do no  longer have identical velocity space axes. A suitable interpolation or reconstruction has to take place in the spatial transport, thus potentially negating the advantage in numerical diffusivity.
\end{itemize}

For the actual process of solving the Vlasov equation, a fundamental decision has to be made in the structure of the code, whether the solution steps are to be performed in a proper 6D manner \citep[e.g.][]{VogmanPhD}, or whether a Strang-splitting approach will be taken \citep{StrangSplitting,ChengVlasov1976,Mangeney2002}, in which the position and velocity space solution steps are performed independently of each other. Due to the large number of involved dimensions, and thus computational context of each solver step, the latter approach tends to have significant performance benefits, whilst still achieving convergence \citep{EinkemmerStrangSplitting}. Alternative time-splitting methods based on Hamiltonians have also been proposed \citep[e.g.][]{Crouseilles2015,Casas2017}

If a Cartesian velocity grid is employed in the simulation, multiple families of solvers are available for it \citep{Filbet2003}. In all cases, the effects of diffusivity of the solver need to be considered. Especially uncontrolled velocity space diffusion manifests itself as numerical heating, as the distribution function tends to broaden over time. Higher orders of solvers and reconstruction methods, as well as explicit formulations in which moments of the distribution function are conserved are therefore advisable \citep{Balsara2017}.

The choice of a Eulerian representation of phase space brings certain implementation details for High Performance Computing (HPC) aspects with it. The relative ease of spatial decomposition into independent computational domains, which communicate through ghost cells, can be employed readily for Eulerian Vlasov simulations, providing a straightforward path to parallel implementations. On the other hand, the inherent limitations of Eulerian schemes (such as conditions for time steps, compare section \ref{sec:CFL}) limit their overall numerical efficiency, and the high-dimensional nature of phase space can lead to challenges in appropriately represented and resolved numerical grids. As so often in HPC, design decisions have to be based on the specific properties of the physical system under investigation.

\subsubsection*{Finite volume solvers}
\label{sec:FVM}

As the Vlasov equation (\ref{eq:vlasov}) is fundamentally a hyperbolic conservation law in 6D, it can be solved using the well-established methods of Finite Volumes \citep[FV,][]{Leveque}. In this approach, the phase-space fluxes are calculated through each interface of a discrete simulation volume (or phase-space cell) by reconstructing the phase-space density distribution through an appropriate interpolation scheme. The characteristic velocities at both sides of this interface are determined and the Riemann problem \citep{toro2014riemann} at each of these interfaces is solved to update the material content in each cell.

If Strang splitting is used to perform separate spatial and velocity-space updates, it is noteworthy that the state vector only contains a single scalar quantity (the phase-space density) and each cell interface update only needs to take a single characteristic velocity into consideration: For the update in a spatial direction, the characteristic is given by the corresponding cell's velocity space coordinates, whereas in the velocity space update step, the acceleration due to magnetic, electric and external field forces is homogeneous throughout each spatial cell. The Riemann problem for the Vlasov update does therefore not require the solution or approximation of an eigenvalue problem, which significantly simplifies its solution in comparison to hydrodynamic or MHD FV solvers. This property also enables the efficient use of semi-Lagrangian solvers (discussed more in section \ref{sec:semilag}).

As will be shown in section \ref{sec:Vlasiator}, versions of the Vlasiator code until 2014 employed a FV formulation of its phase space update \citep{vonAlfthan2014} and numerous other implementations exist \citep{Banks2010,Wettervik2017}. A comprehensive introduction to the implementation and thorough analysis of the behaviour of a fully 6D implementation of a FV Vlasov solver is given by \citet{VogmanPhD}.

\subsubsection*{Finite difference solvers}

While the Vlasov equation \eqref{eq:vlasov} could in principle be solved by directly employing finite difference methods, this approach does not seem to be favoured, and its applications in the literature appear to be limited to fundamental theory studies only \citep[e.g.,][]{Schaeffer1998,Holloway1995}.
The biggest issue with finite difference formulations is the lack of explicit conservation moments of the distribution function and related quantities. While high-order methods can still maintain suitable approximate conservation properties, their computational demands and/or diffusivity make them impractical.

\subsection{Spectral solvers}
\label{subsec:fourierSolvers}

Instead of a direct discretisation of the phase-space density $f(\vec{x},\vec{v}, t)$ in its $\vec{x}$ and $\vec{v}$ coordinates, a change of basis functions can be employed, each coming with benefits and limitations. The transformation of $f$ into a different basis can be performed in the velocity coordinates only (cf. Figure \ref{fig:representations}b), or in both spatial and velocity coordinates, depending on the physical application.

If a splitting scheme is employed, where velocity and real space advection updates are treated separately, the advection in a Fourier-transformed
coordinate can be completely performed in Fourier space, as the transform of any coordinate $x \rightarrow k_x$ results in the differential advection operator $v_x \,\nabla_x$ turning into a simple multiplication:
\begin{equation}
   v_x \, \nabla_x f\left(x\right) \xrightarrow{\mathrm{\small{Fourier}}} \mathrm{i}\, v_x k_x \, \tilde{f}\left(k_x\right).
\end{equation}
However, for the acceleration update, this transformation brings in the additional complication that the acceleration $\vec{a}$ would have to be independent of $\vec{v}$, which is true for the electrostatic Vlasov-Poisson system, but incorrect in full Vlasov-Maxwell scenarios, due to the $v$-dependence of the Lorentz force. In order to accommodate velocity-dependent acceleration, solving a system in such a way typically requires multiple forward and backward Fourier transforms within one time  step \citep{KlimasFarrell1994}.

%

The limit of filamentation in a thus-represented velocity space becomes the question of which maximum velocity space frequency $\vec{k}_{v,\text{max}}$ is available, and the filamentation problem itself becomes a boundary problem at the maximum extents of velocity $\vec{k}$-space \citep{eliassonVlasov}. However, stability issues of this scheme remain under discussion \citep{Figua2000, Klimas2017}.

Finally, a full Fourier-space representation of $\tilde{f}\left(\vec{k}_r,\vec{k}_v, t\right)$, in which also the real space coordinate $\vec{x} \rightarrow \vec{k}_r$ is transformed is a possibility. However, it further complicates the treatment of configuration and velocity space boundaries \citep{eliassonBoundaries}. When used with periodic spatial boundary conditions, such a setup can be quite efficient for the study of kinetic plasma wave interactions.

Apart from the Fourier basis, other orthogonal function systems can be used as the basis for description of phase-space densities. A popular choice is presented by Hermite functions \citep{Delzanno2015, CamporealeHermite2016}, whose $\mathbf{L}^2$ convergence behaviour closely matches that of physical velocity distribution functions, and whose property of being eigenfunctions to the Fourier transform can be used as a numerical advantage.
Since the zeroth Hermite function $H_0(\vec{v})$ is simply a Maxwellian particle distribution, a hybrid-Vlasov code with Hermitian basis functions should replicate MHD behaviour in this limit. Adaptive inclusion of higher-order Hermite functions then allows an increasing amount of kinetic physics to be numerically represented.

A common problem of any kind of spectral method, be it Fourier-based, or using any other choice of nonlocalised basis functions, is the formulation of boundary conditions. While microphysical simulations of wave or scattering behaviour can usually get away with periodic boundary conditions, macroscopic systems will require boundaries at which interaction of plasma with solid or gaseous matter is to be modelled (such as planetary or stellar surfaces), inflow conditions as well as outflow boundaries. Due to the unavailability of suitable spectral formulations for these boundaries, spectral-domain solvers have not gained foothold in modelling efforts of macroscopic astrophysical systems.

In any nonlocal choice of basis function for the phase-space representation, be it Fourier-, Hermite- or wavelet-based \citep{Besse2008}, extra thought has to be put into scalability of parallel solvers for it. If a change of basis function (such as a switch from a real-space to a Fourier space representation) is required as part of the simulation update step, this will typically not scale beyond hundreds of cores in supercomputing environments.

\subsection{Tensor train}

An entirely separate class of numerical representations for the phase-space density is provided by the tensor train formalism \citep{KormannTensorTrain} illustrated in Figure \ref{fig:representations}c. The idea behind this approach is inspired by Strang-splitting solvers, in which spatial and velocity dimensions are treated in individual and subsequent solver steps. The overall distribution function $f(x_1,x_2,\ldots,x_n)$ is represented as a tensor product of component basis functions,
\begin{equation}
	f(x_1,x_2,\ldots,x_n) = \prod_{k=1}^n f_k(x_k)
\end{equation}
in which each $f_k(x_k)$ is only dependent on a single coordinate $x_k$, and thus only affected by a single dimension's update step. The generalised formulation is called the \emph{Tensor Train} of ranks $r_1\ldots r_n$ (compare Fig.\ \ref{fig:representations}c),
\begin{equation}
	f(x_1,x_2,\ldots,x_n) = \sum_{\alpha_1=1}^{r_1} \cdots \sum_{\alpha_n=1}^{r_n} \prod_{k=1}^n Q_k(\alpha_k-1, x_k, \alpha_k),
    \label{eq:tensortrain}
\end{equation}
in which the distribution function is entirely formulated in terms of sums of products of $Q_k$, which themselves only depend on a single coordinate $x_k$. Transport can be performed by individually affecting each $Q_k$, followed by a rounding step to keep the tensor train compact.

While this approach has so far only been employed in low-dimensional approaches and for feasibility studies, and attempts at large numerical simulations using tensor train models have not yet been performed, efforts to integrate them into existing codebases are underway.

\subsection{Semi-Lagrangian and fully Lagrangian solvers}
\label{sec:semilag}

\begin{figure}
	\includegraphics[width=\textwidth]{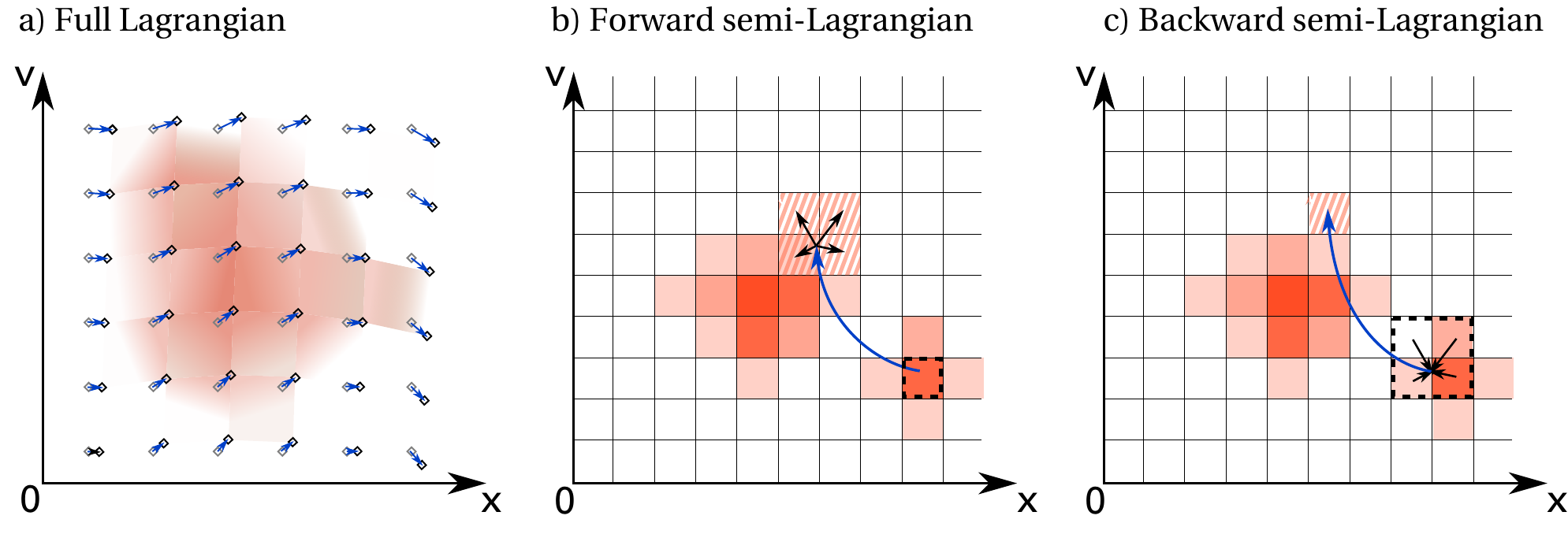}
    \caption{Illustration of Lagrangian and semi-Lagrangian approaches: \textbf{a)} In a full Lagrangian solver, phase-space samples are moved along their characteristic trajectories, never mapped to a grid and the phase-space values in between are obtained by interpolation. \textbf{b)} In a forward semi-Lagrangian method, the same process is used, but the phase-space values are redistributed to a grid at regular intervals. \textbf{c)} A backward semi-Lagrangian method follows the characteristic of every target phase-space point backwards in time and obtains the source value by interpolation.}
        \label{fig:semilag}
\end{figure}
As a consequence of Liouville's theorem (cf. section \ref{sec:liouville}), numerical solutions of the Vlasov equation can be elegantly formulated in Lagrangian and semi-Lagrangian ways, by following the characteristics in phase space. Since the spatial velocity of any point in phase space is simply given by its velocity space coordinates, and its acceleration due to Lorentz' force is provided by the local electromagnetic field quantities, a unique characteristic for each point in phase space is easily obtained in a simulation (cf. section \ref{sec:FVM}). 

As the simulation progresses, the distribution of these sample points will shift, maintaining their initial phase-space density values, and the volumes in between them obtain phase-space density values through interpolation. If necessary, new samples can be created, or existing ones merged, where filamentation requires it.
In essence, fully Lagrangian simulation codes \citep{Kazeminezhad2003,Nunn2005,Jenab2014} track the motion of samples of density through phase space, stepping forward in time, resulting in an updated phase-space distribution. This is illustrated in Figure \ref{fig:semilag}a. Sometimes, these methods are referred to as Lagrangian particle methods, as each phase-space sample can be modelled as a macroparticle.

Much more common than the fully Lagrangian formulation of Vlasov solvers is the family of semi-Lagrangian solvers \citep{Sonnendrucker1999}. In these, the phase-space samples to be propagated are obtained at every time step from a Eulerian description of phase space, their transport along the characteristics is calculated within the time step, and the resulting updated phase-space density is sampled back into a Eulerian grid \citep[which can be either structured or unstructured, see][]{Besse2003}. This process can be performed either forwards in time \citep[][see Figure \ref{fig:semilag}b]{Crouseilles2009}, in which the source grid points are scattered into the target locations, or backwards in time \citep{Sonnendrucker1999,PfauKempf2016PhD}, where each target grid point performs a gather operation, spatially interpolating inside the previous state of the time steps (Fig.\ \ref{fig:semilag}c). Backwards semi-Lagrangian methods are sometimes also referred to as Flux Balance Methods \citep[see][]{Filbet2001}. Either way, an interpolation step will be involved which may again lead to significant numerical diffusion, unless methods are used to minimise it. Some of the more common interpolation procedures used are cubic splines and Hermite reconstruction because they produce smooth results with reasonable accuracy and are less dissipative than other methods using continuous interpolations \citep{Filbet2003b}. Lagrange interpolation methods produce more accurate results but require higher order polynomials and large stencils to limit diffusion. The high-order discontinuous Galerkin method for spatial discretisation, along with a semi-Lagrangian time stepping method, has also been used in Vlasov-Poisson systems providing an improvement in accuracy compared to previously used techniques \citep{Rossmanith2011}. The flexibility of combining different approaches is also seen in a recent particle-based semi-Lagrangian method for solving the Vlasov-Poisson equation \citep{Cottet2018}

\citet{ChengVlasov1976} were the first authors to employ semi-Lagrangian updates of a Vlasov-Poisson problem in a Strang-splitting setup, which they refer to as the time-splitting scheme, in which they independently treated advection due to temporal and spatial updates. \citet{Mangeney2002} later formulated a Strang-splitting scheme for the Vlasov-Maxwell equation.
As such a splitting scheme performs acceleration and translation steps separately, the phase-space trajectories of any simulation point approximates their physical behaviour in a staircase-like dimension-by-dimension pattern.






\subsection{Field solvers} \label{sec:fs}

The Vlasov equation does not stand alone in describing the physical system in consideration, but requires a further prescription of the fields introducing the force terms. In the vast majority of cases in computational astrophysics, these will be electromagnetic forces self-consistently produced through the motion of charged particles in plasma, although there have been examples of Vlasov-gravity simulations \citep{Guo2008}, in which the Poisson equation was solved based on the simulation's mass distribution. Also in a few cases, the fields affecting phase-space distributions are considered entirely an external simulation input, with no feedback from the phase-space density onto the fields \citep{Palmroth2013}, which can be called ``test-Vlasov'' simulations, in analogy to test-particle runs. These are particularly useful as test cases before the fully operational code can be launched.

A key requirement for any field solver is to preserve the solenoidality of the magnetic field $\mathbf{B}$ expressed by equation~\eqref{eq:Gauss}. There are two main avenues used to achieve this goal \citep[e.g.,][and references therein]{Toth2000,BalsaraKim2004,ZhangFeng2016}. Either the field reconstruction is divergence-free by design, such as the one used in Vlasiator (see section \ref{txt:Vlasiator_solvers}), or a procedure is needed to periodically clean up the divergence of $\mathbf{B}$ arising from numerical errors.

In the following sections, different solvers for electromagnetic fields (and their simplifications) will be discussed in relation to astrophysical Vlasov simulations. These are fundamentally very similar in structure to the field solvers used in other simulation methods, such as PIC and MHD, and can in many cases be adapted directly from these with little change.

\subsubsection*{Electrostatic solvers}

If modelling a physical system in which the interaction of plasma with magnetic fields is of little importance (such as electrostatic wave instabilities, dusty plasmas, surface interactions \citep{ChaneYook2006} and other typically local phenomena), the magnetic field force ($q \vec{v} \times \vec{B}$) part of the Vlasov equation can be neglected, and a purely electrostatic system remains.

Neglecting the effects of $\vec{B}$ completely leads to the Vlasov-Poisson system of equations \eqref{eq:electrostaticVlasov} and \eqref{eq:Poisson}, for which the field solver needs to find a solution to the Poisson equation at every time step. Being an elliptic differential equation that is solved instantaneous in time, no time step limit arises from the field solver itself.
Typically, solvers use approximate iterative approaches, multigrid methods or Fourier-space solutions \citep{Birdsall2004}.

Another option, if an initial solution for the electric field has been found (or happens to be trivial), is to update it in time by using Ampere's equation in the absence of $\vec{B}$,
\begin{equation}
\frac{\partial \mathbf E}{\partial t} = - \frac{\mathbf j}{\epsilon_0}
\end{equation}
in either an explicit finite-difference manner, or using more advanced implicit formulations \citep{CHENG2014}. Special care should however be taken to prevent the violation of the Gauss law (cf. equation~\eqref{eq:Gauss}) by using appropriate numerical methods.

\subsubsection*{Full electromagnetic solvers}

If the full plasma microphysics of both electrons and ions is to be considered, and particularly if radio wave or synchrotron emissions are intended outcomes of the system, one must use the full set of Maxwell's equations. A popular and well-established family of electromagnetic field solvers is the finite difference time-domain (FDTD) approach, which has a longstanding history in electrical engineering applications. In formulating the finite differences for the $\partial \vec{E} / \partial t \sim \nabla \times \vec{B}$ and $\partial \vec{B} / \partial t \sim \nabla \times \vec{E}$ terms of Maxwell's equations, it is often advantageous to use a staggered-grid approach, in which the electric field and magnetic field grids are offset from one another by half a grid spacing in every direction \citep{Yee1966}. In this setup, every component of the electric field vector is surrounded by magnetic field components and vice versa, so that the finite difference evaluation of the rotation can be performed without any need for interpolation.

Care should be taken when employing FDTD solvers for studies of wave propagation at high frequencies or wave numbers, as the numerical dispersion relations of waves are deviating from their physical counterparts for high $\vec{k}$, and this effect in particular is anisotropic in nature, and most strongly pronounced in cases of diagonal propagation, due to the intrinsic differences in the manner by which grid-aligned and non-grid-aligned computations are handled. \citet{Karkkainen2006} and \citet{VayCole2011} present a thorough analysis of this problem in the case of PIC simulations, and provide suggestions for mitigating their effects. The largest disadvantage of FDTD solvers is their stringent requirement to resolve the propagation of fields at the speed of light, thus leading to extremely short time step lengths. In order to simulate anything at non-microscopic timescales, other methods will need to be used. Fourier-space solvers of Maxwell's equations are advantageous in this respect, as they do not come with fundamental time step limitations. This is weighed up by the fact that their parallelisation is more difficult, and the formulation of appropriate boundary conditions is not always possible (cf. section \ref{subsec:fourierSolvers}).

\subsubsection*{Hybrid solvers}

If large-scale phenomena with timescales much larger than the local light crossing time are being investigated, FDTD Maxwell solvers quickly lose their appeal. If magnetic field phenomena are still to be considered self-consistently in the simulation, appropriate modifications of the electrodynamic behaviour have to be taken, so that their simulation with longer time steps becomes feasible.

One common way to get rid of the speed of light as a limiting factor is by getting rid of the electromagnetic mode as a solution to Maxwell's equation altogether, in a process called the \emph{Darwin approximation} \citep[see section \ref{txt:hybrid_vlasov} and][]{Schmitz2006, Bauer2005}. In this process, the electric field is decomposed into its longitudinal and transverse components $\vec{E} = \vec{E}_L + \vec{E}_T$, with $\nabla \times \vec{E}_L = 0$ and $\nabla \cdot \vec{E}_T = 0$. Only $E_L$ is allowed to participate in the temporal update of $\vec{B}$, so that the electromagnetic mode drops out of the simulated physical system. As a result, the fastest remaining wavemode in the system becomes the Alfv\'en wave, and the maximum time step rises significantly, by a factor of $c/v_A$.

Approximating the full set of Maxwell equations comes at the cost of not having a closed set of equations any more. As already shown in section \ref{txt:hybrid_vlasov}, the system is typically closed by providing a relation between the electric and magnetic field such as equation~\eqref{eq:Ohm}, called Ohm's law. The level of complexity of Ohm's law directly influences the simulation results as it immediately affects the kinetic physics described by the model.

\subsection{Coupling schemes}
\label{sec:couplingschemes}
A reduction of the computational burden of a model can be achieved by coupling different schemes in order to focus the use of the costlier kinetic model on the region(s) of interest while solving other parts of the system with less intensive algorithms. This is also a means of extending the simulation domain where one system is taken as the boundary condition of the other. Various classes of coupled models exist, depending on the coupling interface chosen.

One strategy is to define a spatial region of interest in which the expensive kinetic model is applied, embedded in a wider domain covered by a significantly cheaper fluid model. While the method is under investigation and has been tested on classic small problems \citep{Rieke2014JComputPhys}, it has not been applied in the context of large-scale astrophysical simulations yet. However, this type of coupling is being used successfully in the case of fluid-PIC coupling \citep[e.g.][]{Toth2016EPICMHD,Chen2017EPICMHD} and also in reconnection studies \citep[e.g.][]{Usami2013}. The disadvantage of this strategy is that scale coupling cannot be addressed as the kinetic effects do not spread into the fluid regime, and smaller-scale physics can only affect the solution in the domain at which the kinetic physics is in force.

Another strategy consists in defining the regions of interest in velocity space, that is coupling a fluid scheme describing the large-scale behaviour of a system with a Vlasov model handling suprathermal populations introducing kinetics into the model. Again, this is a recent development for which a certain amount of theoretical work and testing on small cases has been done \citep[e.g.][]{Tronci2014PPCF} but not yet extended to larger scale applications.

\subsection{Computational burden of Vlasov simulations}
\label{txt:computational_burden}
Representing numerically the complete velocity phase space of a kinetic plasma system including all required physical processes is computationally intensive, and a large amount of data needs to be stored and processed. Different possible representations of the phase-space distribution and solution methods and their expected scaling shall be given in this section. Computational requirements for equivalent PIC simulations are estimated in comparison, although due to their different tuneable parameters, a rigorous comparison is difficult and beyond the scope of this work.

As shown in section \ref{sec:eulerian}, a blunt Eulerian discretisation of a magnetospheric simulation without any velocity space sparsity results in $10^{18}$ sample points or a minimum of 4\,EiB memory requirement, which is unrealistic on current and next-generation architectures. A first approach is to reduce the dimensionality from a full 3D space to a 2D slice, which results in a reduction of sample points of the order of $10^4$. Obviously a further reduction to 1D yields a similar gain. With a sparse velocity space strategy as used in Vlasiator (see Section \ref{sec:vlasiatorSparse} below) a further reduction by a factor of $10^2-10^3$ sample points can be achieved. Typically modern large-scale kinetic simulations both with Vlasov-based methods \citep[e.g.][]{Palmroth2017AnGeo} and PIC methods \citep[e.g.][]{Daughton11} reach an order of magnitude of $10^{11}-10^{12}$ sample points.

Beyond the number of sample points to be treated, the length of the propagation time step relative to the total simulation time aimed for is a crucial component of the computational burden of a model. Certain classes of solvers are limited in that respect as they cannot allow a signal to propagate more than one sampling interval or discretisation cell per time step (see section\ \ref{sec:CFL}). With respect to hybrid models using the Darwin approximation, the inclusion of electromagnetic (light) waves in the model description results in a reduction of the allowable time step by a factor of $10^3$ or more. Eulerian solvers typically have similar limitations which can impact the time step by a factor of approximately $10^2$ due to the Larmor motion in velocity space in the presence of magnetic field. Subcycling strategies and the use of Lagrangian algorithms are common approaches to alleviate these issues, at the potential cost of some physical detail however.

A choice of basis function for the representation of velocity space other than Eulerian grids (like spectral or Hermite bases) can in many cases be beneficial to limit the memory requirements for reasonable approximations of the velocity space morphology. Care must however be taken that non-local transformations from one basis to another, such as a Fourier transform, tend to have unfavourable scaling behaviour in massively parallel implementations.
Tensor-train formulations appear to be a promising avenue for the representation of phase space densities that have suitable computational properties, but large-scale space plasma applications have not been demonstrated yet.

Higher computational requirements are expected if physics of multiple particle species (especially electrons) are essential for the system under investigation. The need to represent multiple separate distribution functions multiplies the memory and computation requirements. The relative mass ratios of these species also have an effect on the kinetic time and length scales that need to be resolved. Going from a purely proton-based hybrid-Vlasov to a ``full-Vlasov'' simulation, in which electrons are included as a kinetic species shortens the Larmor times by a factor of $m_p / m_e = 1836$ and depending on the employed solver may require resolution of the plasma's Debye length. The latter factor means that, with respect to the reference hybrid simulation considered above, which approximately resolves the ion kinetic scales, a spatial resolution increase of the rough order of $10^5$ would be required (see Table \ref{tab:Scales}), amounting to a staggering $10^{15}$ more sampling points. In order to reduce this considerable overhead, a common approach is to rescale physical quantities such as the electron-to-proton mass ratio and/or the speed of light \citep[e.g.][]{HockneyEastwood}, while hoping to maintain quantitatively correct kinetic physics behaviour.

Most of these scaling relations likewise apply in PIC. In these, however, the parameter most strongly affecting the computational burden of the phase space representation is the particle count. As a rule of thumb, a PIC simulation with a particle count similar to the sampling point count of an equivalent Vlasov simulation will have a similar overall computational cost. For many physical scenarios, this particle count can be chosen to be significantly lower (on the order of 100-1000 particles/cell), especially if noisy representations of the velocity spaces are acceptable. 
In simulations with high dynamic density contrasts, in which certain simulation regions deplete of particles, as well as setups in which a minimisation of sampling noise is essential (such as investigations of nonlinear wave phenomena), PIC and Vlasov simulations are expected to reach a break-even point.

\subsection{Achievements in Vlasov-based modelling}
The progress of available scientific computing capabilities towards and beyond petascale in the last decade has driven the interest in and applicability of Vlasov-based methods to multidimensional space and astrophysical plasma problems. Table \ref{tab:Applications} compiles existing research work using direct solutions of the Vlasov equation in plasma physics. Table \ref{tab:Applications} only includes works with a direct link to space physics and astrophysics, meaning that purely theoretical work as well as research from adjacent fields, in particular nuclear fusion and laser-plasma interaction, has been omitted from this list on purpose. As of 2018, the Vlasov equation has thus been used in space plasma physics and plasma astrophysics to model magnetic reconnection, instabilities and turbulence, the interaction of the solar wind with the Earth and other bodies, radio emissions in near-Earth space and the charge distribution around spacecraft.

\begin{table}
\centering
   \begin{tabularx}{\textwidth}{X|X|X}
      \hline
      Application & Model characteristics & References \\
      \hline
      \hline
      Magnetic reconnection & SL 2D-3V full Vlasov & \citet{Umeda2009,Umeda2010,Zenitani2014} \\
      & FD 2D-3V hybrid-Vlasov (H\textsuperscript{+}) & \citet{Cerri2017R,Franci2017} \\
      \hline
      Kelvin-Helmholtz instability & SL 2D-3V full Vlasov & \citet{Umeda2010b,Umeda2014KH} \\ 
      \hline
      Rayleigh-Taylor instability & SL 2D-2V full Vlasov & \citet{Umeda2016RT,Umeda2017RT} \\
      \hline
      Solar wind turbulence & FD 3D-3V hybrid-Vlasov (H\textsuperscript{+}) & \citet{Cerri2017a,Servidio2015} \\
	  & FD 2D-3V hybrid-Vlasov (H\textsuperscript{+}) & \citet{Cerri2016,Cerri2017b,Leonardis2016,Pucci2016,Servidio2012,Servidio2014,Valentini2010,Valentini2011,Valentini2014,Valentini2016,Vasconez2014,Vasconez2015} \\
                            & FD 2D-3V hybrid-Vlasov (H\textsuperscript{+}, He\textsuperscript{++}) & \citet{Perrone2012,Perrone2014,Perrone2014b} \\
      \hline
VLF radio emissions in the Earth's radiation belt & L 1D-3V hybrid-Vlasov (e$^-$) & \citet{Harid2014,Nunn1997,Nunn2005} \\
      & SL 1D-3V hybrid-Vlasov (e$^-$) & \citet{Gibby2008} \\
      \hline
      Solar wind interaction with unmagnetised or weakly magnetised bodies & SL 2D-3V full Vlasov & \citet{Umeda2011,Umeda2012,Umeda2013,Umeda2014,Umeda2015} \\
\hline
      Solar wind interaction with the terrestrial magnetosphere & FV 3D-3V test-Vlasov (H\textsuperscript{+}) & \citet{Palmroth2013} \\

& FV 2D-3V hybrid-Vlasov (H\textsuperscript{+}), equatorial plane & \citet{Pokhotelov2013,vonAlfthan2014,Kempf2015} \\

& SL 2D-3V hybrid-Vlasov (H\textsuperscript{+}), polar and equatorial plane & \citet{Palmroth2015,PfauKempf2016,Hoilijoki2016,Hoilijoki2017,Palmroth2017AnGeo} \\
      \hline
      Charge and potential distribution around a spacecraft & 3D Vlasov-Poisson (iterative relaxation algorithm) and Vlasov-Laplace (Lagrangian) & \citet{ChaneYook2006} \\
      \hline
      Relativistic Weibel instabilities & SL 1D-2V hybrid-Vlasov (e$^-$) & \citet{Inglebert2011} \\
        & SL 2D-2V hybrid-Vlasov (e$^-$) & \citet{Ghizzo2017} \\
        & SL 2D-2V and 2D-3V hybrid-Vlasov (e$^-$) & \citet{Sarrat2017} \\
      \hline
      \hline
   \end{tabularx}
   \caption{Space and astrophysical applications of Vlasov-based plasma simulation methods. FD: Finite Difference. FV: Finite Volume. L: Fully Lagrangian. SL: Semi-Lagrangian. e\textsuperscript{-}, H\textsuperscript{+}, He\textsuperscript{++}: kinetic species (electrons, protons, helium ions) in a hybrid setup.}
\label{tab:Applications}
\end{table}

\section{Vlasiator}\label{sec:Vlasiator}

This section considers the choices and approaches made for the Vlasiator code, attempting to describe the near-Earth space at ion kinetic scales. 
Vlasiator simulates the global near-Earth plasma environment through a hybrid-Vlasov approach. The evolution of the phase-space density of kinetic ions is solved with Vlasov's equation (Eq. \ref{eq:vlasov}), with the evolution of electromagnetic fields described through Faraday's law (Eq. \ref{eq:Faraday}), Gauss' law and the solenoid condition (Eq. \ref{eq:Gauss}), and the Darwin approximation of Amp\`ere's law (Eq. \ref{eq:AmpereDarwin}). Electrons are modelled as a massless charge-neutralising fluid. Closure is provided via the generalised Ohm's law (Eq. \ref{eq:Ohm}) under the assumptions of high conductivity, slow temporal variations, and cold electrons, i.e. the Hall MHD Ohm's law (Eq. \ref{eq:HallMHD}). 
The source code of Vlasiator is available at \texttt{http://github.com/fmihpc/vlasiator} according to the Rules of the Road mapped out at \texttt{http://www.physics.helsinki.fi/vlasiator}.

\subsection{Background}

Vlasiator has its roots in the discussions within the global MHD simulation community around 2005. It was becoming evident that while global MHD simulations are important, their capabilities, especially in the inner magnetosphere, are limited. The inner magnetosphere consists of spatially overlapping plasma populations of different temperatures \citep[e.g.][]{Baker95} and therefore the environment cannot be satisfactorily modelled with MHD to a degree allowing e.g., environmental predictions for societally critical spacecraft or as a context for upcoming missions, like the Van Allen Probes \citep[e.g.][]{fox13}. To this end, two strategies emerged, including either coupling a global MHD simulation with an inner magnetospheric simulation \citep[e.g.][]{huang06}, or going beyond MHD with the then newly introduced hybrid-PIC approach \citep[e.g.][]{omidi07}. Coupling different codes carries a risk that the effects of the coupling scheme dominate over the improved physics. On the other hand, while hybrid-PIC simulations had produced important breakthroughs \citep[e.g.][]{omidi05}, the velocity distributions computed through binning are noisy due to the limited number of launched particles, which could compromise physical conclusions. Further, due to the limited number of particles, the hybrid-PIC simulations could not provide sharp gradients, which would become a problem especially in the magnetotail, where the lobes surrounding the dense plasma sheet are almost empty. As the tail physics is critical in the global description of the magnetosphere, the idea about a hybrid-Vlasov simulation emerged.

The objective was simple, just to go beyond MHD by introducing protons as a kinetic population modelled by a distribution function and thus getting rid of the noise. Several challenges were identified. First, if one neglects electrons as a kinetic population, one will, e.g., lose electron-scale instabilities that can be important in the tail physics \citep[e.g.,][]{pritchett05}. Second, a global hybrid-Vlasov approach is still an extreme computational challenge even with a coarse ion-scale resolution, since it must be carried out in six dimensions. Further, doubts existed about whether grid resolutions achievable with current computational resources would facilitate ion kinetic physics. However, with a new approach without historical heritage, one could utilize the latest high-performance computing methods and new computational architectures, provided that the code would always be portable to the latest technology. The computational resources were still obeying the ``Moore law'', and petascale systems had just become operational \citep{kogge09}. With these prospects in mind, Vlasiator was proposed to the newly established European Research Council in 2007, which solicited new ideas with a high risk -- high gain vision.

\subsection{Grid discretisation} 
\label{sec:griddiscretation}
\label{sec:vlasiatorSparse}

The position space is discretised on a uniform Cartesian grid of cubic cells. Each cell holds the values of variables that are either being propagated or reconstructed (e.g.\ the electric and magnetic fields and the ion velocity distribution moments) as well as housekeeping variables. In addition, a three-dimensional uniform Cartesian velocity space grid is stored in each spatial cell. For position space Vlasiator uses the Distributed Cartesian Cell-Refinable Grid library \citep[DCCRG;][]{Honkonen2013CPC} albeit without making use of the adaptive mesh refinement capabilities it offers. The library can distribute the grid over a large supercomputer using the domain decomposition approach (see section \ref{txt:Vlasiator_parallelisation} for details on the parallelisation strategies).

\begin{figure}
   \centering
   \includegraphics[width=\textwidth]{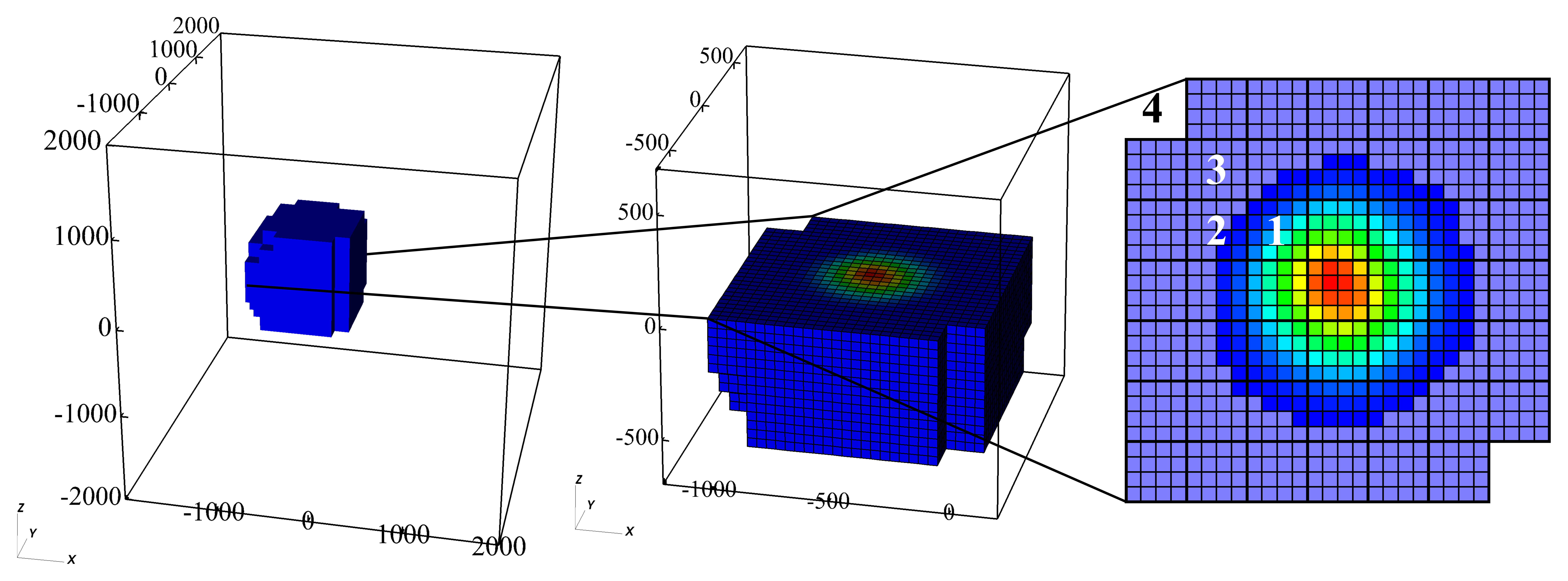}
   \caption{Illustration of the sparse velocity space. Left: full extent of velocity space including a population drifting at $V_x\approx-500\,\mathrm{km/s}$. Middle: Cut through the population. 
Right: Slice showing the cells with $f>f_\mathrm{min}$ in full colour and retained neighbours greyed out. 1: block fully above $f_\mathrm{min}$. 2: block partially above $f_\mathrm{min}$. 3: retained 
neighbouring block. 4: disregarded block without neighbours above $f_\mathrm{min}$. See section \ref{txt:Vlasiator_parallelisation} for details on the block structure. Figure from \citet{PfauKempf2016PhD}.}
   \label{fig:sparsity}
\end{figure}

The velocity space grid is purpose built for that specific task. A major performance gain in terms of memory and computation is achieved by storing and propagating the volume average of $f$ in every cell at position $\mathbf{x}$ only in those velocity space cells where $f$ exceeds a given density threshold $f_\mathrm{min}$. In order to accurately model propagation and acceleration, a buffer layer is maintained by modelling also cells that are adjacent in position or velocity space. The principle is illustrated in Figure \ref{fig:sparsity}. This threshold can be constant or scaled linearly with the ion density. For each ion population, the maximal velocity space extents and the resolution are set by the user. This so-called sparse velocity space strategy allows to increase the resolution and track the distribution function in the regions where it is present instead of wasting computational resources covering the full extents of reachable velocity space. It is however important to set the value of $f_\mathrm{min}$ carefully in order to conserve the moments of $f$ (density, pressure, etc.) to the desired accuracy and in order to include in a given simulation all expected features such as low-density high-energy populations. A detailed discussion of the effects of the grid discretisation parameters on the simulation of a collisionless shock was published by \citet{PfauKempf2018}.

\subsection{Solvers and time-integration}
\label{txt:Vlasiator_solvers}
The structure of the hybrid-Vlasov set of equations leads to the logical split into a solver for the Vlasov equation and a solver for the electric and magnetic field propagation.

\subsubsection*{Vlasov solver}

In advancing the Vlasov equation, Vlasiator utilises Strang splitting \citep[][and references therein]{Umeda2009,Umeda2011}, where updates of the particle distribution functions are performed, separately using a spatial translation operator $S_T$ for advection; $S_T$ = (${\bf v}\cdot\frac{\partial f_\mathrm{s}}{\partial {\bf x}}$), and an acceleration operator $S_A$ = $(\frac{q_\mathrm{s}}{m_\mathrm{s}}{\bf E}\cdot\frac{\partial f_\mathrm{s}}{\partial {\bf v}})$ including rotation $(\frac{q_\mathrm{s}}{m_\mathrm{s}}({\bf v}\times{\bf B})\cdot\frac{\partial f_\mathrm{s}}{\partial {\bf v}})$ for each phase-space volume average. The splitting is performed using a standard leapfrog scheme, where
\begin{equation*}
\widetilde{f}^{N+1} = S_A(\tfrac{\Delta t}{2}) S_T(\Delta t) S_A(\tfrac{\Delta t}{2}) \widetilde{f}^{N}.
\end{equation*}
Acceleration over step length $\Delta t$ is thus calculated based on the field values computed at the mid-point of each acceleration step, i.e. at each actual time step as used for translation.

A global time step is defined, with time advancement calculated for distribution functions and fields in separate yet linked computations. Earlier versions of Vlasiator have used finite volume (FV) Vlasov solvers. In the earliest versions of the code, a FV method based on the solver proposed by \citet{Kurganov2000} was used \citep{Palmroth2013}. A Riemann-type FV solver \citep{Leveque1997,Langseth2000} was used in subsequent works \citep{Kempf2013,Pokhotelov2013,Sandroos2013,vonAlfthan2014,Sandroos2015,Kempf2015}. For these solvers \label{sec:CFL}the classical Courant--Friedrichs--Lewy (CFL) condition \citep{Courant1928} for maximal allowable time steps when calculating fluxes from one phase-space cell to another is
\begin{equation*}
\Delta t < \mathrm{min} \left( \frac{\Delta x_i}{\mathrm{max}(|v_i|)}, \frac{\Delta v_i}{\mathrm{max}(|a_i|)} \right),
\end{equation*}
where $i$ is indexed over three dimensions. In previous versions the CFL condition was found to be very limiting. 
Vlasiator utilizes a semi-Lagrangian scheme \citep[SLICE-3D,][]{Zerroukat2012}\label{sec:SLICE3D}, in which mass conservation is ensured by a conservative remapping from a Eulerian to a Lagrangian grid. Note however that the sparse velocity space strategy implemented in Vlasiator (see section \ref{sec:vlasiatorSparse}) breaks the mass conservation \citep[see][for a discussion of the effect of the phase space density threshold on mass conservation]{PfauKempf2018}. The specificity of the SLICE-3D scheme is to split the full 3D remapping into successive 1D remappings, which reduces the computational cost of the spatial translation and facilitates its parallel implementation.

The velocity space update due to acceleration $S_A(\frac{\Delta t}{2})$ will generally be described by an offset 3D rotation matrix (due to gyration around $\vec{B}$). As every offset rotation matrix can be decomposed into three shear matrices $S = S_x S_y S_z$, each performing an axis-parallel shear into one spatial dimension \citep{CHEN2000308}, the numerically efficient semi-Lagrangian acceleration update using the SLICE-3D approach is possible: before each shear transformation, the velocity space is rearranged into a single-cell column format parallel to the shear direction in memory, each of which requires only a one-dimensional remapping with a high reconstruction order (in Vlasiator, 5\textsuperscript{th} order reconstruction is typically employed for this step). These column updates are optimised to make full use of vector instructions.
This update method comes with a maximum rotation angle limit due to the shear decomposition of about $22^\circ$, which imposes a further time step limit. For larger rotational angles per time step (caused by stronger magnetic fields), the acceleration can be subcycled.

The position space update $S_T(\Delta t)$ will generally be described by a translation matrix with no rotation, and the same SLICE-3D approach lends itself to it in a similar vein as for velocity space. The main difference is the typical use of 3\textsuperscript{rd} order reconstruction in order to keep the stencil width at two. The use of a semi-Lagrangian scheme allow the implementation of a time step limit
\begin{equation*}
\Delta t < \mathrm{min} \left( \frac{\Delta x_i}{\mathrm{max}(|v_i|)} \right) 
\end{equation*}
based on spatial translation only. This condition constrains the spatial translation of any volume averages to a maximum value of $\Delta x_i$ in direction $i$, accounting for only those velocities within phase-space which have populated active cells (see the sparse grid implementation, section \ref{sec:griddiscretation}). This is employed not due to stability requirements, but rather to decrease communication bandwidth by ensuring that a single ghost cell in each spatial direction is sufficient.

\subsubsection*{Field solver}
The field solver in Vlasiator \citep{vonAlfthan2014} is based on the upwind constrained transport algorithm by \citet{Londrillo2004} and uses divergence-free reconstruction of the magnetic fields \citep{Balsara2009}. It utilizes a second-order Runge-Kutta algorithm including the interpolation method demonstrated by \citet{Valentini2007} to obtain the intermediate moments of $f$ needed to update the electric and magnetic fields \citep{vonAlfthan2014}.

The algorithm is subject to a CFL condition such that the fastest-propagating wave mode cannot travel more than half a spatial cell per time step. As the field solver was extended to include the Hall term in Ohm's law, the CFL limit severely impacts the time step of the whole propagation in regions of high magnetic field strength or low plasma density. If the imbalance between the time step limits from the Vlasov solver and from the field solver is too strong, the computation of the electric and magnetic fields is subcycled such as to retain an acceptable global time step length \citep{PfauKempf2016PhD}.

\subsubsection*{Time stepping}
\begin{figure}
   \centering
   \includegraphics[width=\textwidth]{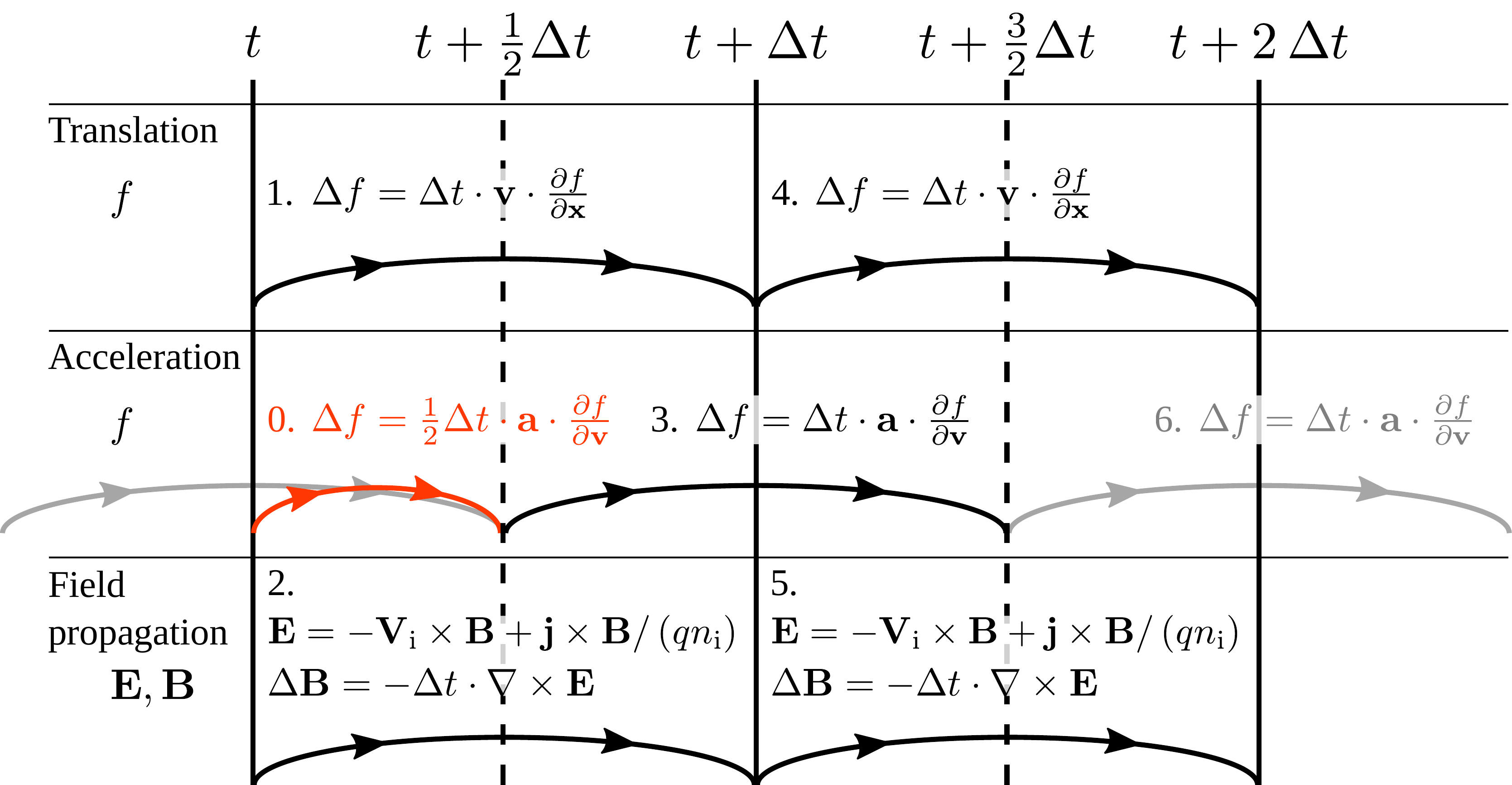}
   \caption{Time stepping in Vlasiator. The translation and acceleration of $f$ are leapfrogged following the Strang-splitting method. The algorithm is initialised by half a time step of acceleration (step 0.\ in red). Then 1.\ $f$ is translated forward by one step $\Delta t$ (possibly subcycled, see text). 2.\ $\mathbf{E},\mathbf{B}$ are stepped forward by $\Delta t$ (possibly subcycled, see text). 3.\ $f$ is accelerated forward by $\Delta t$. The sequence is repeated (4.--6.).}
\label{fig:timestepping}
\end{figure}

The leapfrog scheme of the propagation is initialised by half a time step of acceleration. If the time step needs to change during the simulation due to the dynamics of the system, $f$ is accelerated backwards by half an old time step and forwards again by half a new time step and the algorithm resumes with the new global time step. The complete sequence of the time propagation in Vlasiator is depicted in Figure~\ref{fig:timestepping}, including a synthetic version of the equations used in the different parts.

\subsection{Boundary and initial conditions}\label{sec:boundary_conditions}

	The reference frame used in Vlasiator is defined as follows: the origin is located at the centre of the Earth, the $x$ axis points towards the Sun, the $z$ axis points northward perpendicular to both the $x$ axis and the ecliptic plane, and the $y$ axis completes the right-handed set. This coordinate system is equivalent to the Geocentric Solar Ecliptic (GSE) frame which is commonly used when studying the near-Earth space.
    
    The solar wind enters the simulation domain at the $+x$ boundary, while copy conditions (i.e., homogeneous Neumann boundary conditions obtained by copying the velocity distribution functions and magnetic fields from the boundary cells to their neighbouring ghost cells) are used for the $-x$ boundary and for the boundaries perpendicular to the flow. In the current 2D-3V runs, periodic conditions are applied in the out-of-plane direction (i.e. $+z$ and $-z$ for the ecliptic runs and $+y$ and $-y$ for the polar runs). Currently, three versions of the copy conditions are implemented, which can be adjusted in order to mitigate issues such as self-replicating phenomena at the boundaries. At the beginning of the run or at a restart, the outflow condition can be set to a classic copy condition, to a copy condition where the value of $f$ is modified in order to avoid self-replication or inflowing features, or static conditions can be maintained at the boundary.
    
   The simulation also requires an inner boundary around the Earth, in order to screen the origin of the terrestrial dipole. In the inner magnetosphere, the magnetic field strength increases dramatically, resulting in very small time steps, which would significantly slow down the whole simulation. Also, close to the Earth, the ionospheric plasma can no longer be described as a collisionless and fully ionised medium, and another treatment would be required in order to properly simulate this region. The inner boundary is therefore located at $30,000$ km (about $4.7 \, \mathrm{R}_{\mathrm{E}}$) from the Earth's centre and is currently modelled as a perfect conductor. The distribution functions in the boundary cells retain their initial Maxwellian distributions throughout the run. The electric field is set to zero in the layer of boundary cells closest to the origin, and the magnetic field component tangential to the boundary is fixed to the value given by the Earth's dipole field. Since the ionospheric boundary is given in the Cartesian coordinate system, it is not exactly spherical but staircase-like, introducing several computational problems \citep[e.g.,][]{Cangellaris1991}. This has not been a large problem in Vlasiator up to date, possibly since the computations are carried out in 2D-3V setup. Once the computations are carried out in 3D-3V, this may pose a larger problem, because the magnetic field will  be stronger in 3D near poles.
   
In addition to defining boundary conditions, the phase-space cells  within the simulation box must be initialised to some reasonable values, after which the magnetic and gas pressures and flow conditions cause the state of the simulation to change, and to converge towards a valid description of the magnetospheric system as the box is flushing. The usual method employed in Vlasiator is to initialise the velocity space in each cell within the simulation (excluding the region within the inner boundary) to match values picked from a Maxwellian distribution in agreement with the inflow boundary solar wind density, temperature, and bulk flow direction. The inner boundary is initialised with a constant proton temperature and number density with no bulk flow. The initial phase space density sampling can be improved by averaging over multiple densities obtained via equally spaced velocity vectors within the single velocity space cell.

The Earth's magnetic field closely resembles that of a magnetic dipole, and within the scope of Vlasiator the dipole has been approximated as being aligned with the $z$-axis. For ecliptic $x-y$ plane simulations the dipole field can be used as-is, but for polar $x-z$ plane simulations, a 2D line dipole (which scales as $r^{-2}$ rather than $r^{-3}$) must be used instead in order to prevent the occurrence of unphysical currents due to out-of-plane field curvature. When using this approach, one must calculate the dipole strength that represents reality in some chosen manner, and this is achieved by choosing a value that in turn reproduces the magnetopause at the realistic $\sim10\,\mathrm{R}_\mathrm{E}$ standoff distance (a similar treatment as found in, e.g., \citealt{Daldorff2014}). As the dipole magnetic field is not included in the inflow boundary, there cannot exist a boundary-perpendicular magnetic field component in order to respect the solenoid condition. For ecliptic runs, the dipole field component is aligned with $z$ and thus there is no component perpendicular to the inflow boundary. For polar runs, the dipole field component perpendicular to the inflow boundary must be removed to prevent magnetic field divergence. This is achieved by placing a mirror dipole identical to the Earth's dipole model at the position $(2 \cdot (X_1 - \Delta x),0,0)$ that is at twice the distance from the origin to the edge of the final non-boundary simulation cell. For each simulation cell, the static background magnetic flux through each face is thus assigned as a combination of flux calculated from the chosen dipole field model, a mirror dipole if present, and the solar wind IMF vector. This background field, which is curl-free and divergence-free, is left static, and instead any calculations involving magnetic fields operate on a separate field which acts as a perturbation from this initial field. 

\subsection{Parallelisation strategies}
\label{txt:Vlasiator_parallelisation}

Given the curse of dimensionality in Vlasov simulations (cf. section \ref{sec:dimensionality}), the amounts of memory and computational steps required for global magnetospheric hybrid-Vlasov simulations are extreme. Therefore, the use of supercomputer resources and parallelisation techniques is essential. Vlasiator uses three levels of parallelisation, of which the first employed is the decomposition of the spatial simulation domain into subdomains handling individual tasks using the Message-Passing Interface \citep[MPI,][]{MPI3.1}. The use of the DCCRG grid library \citep{Honkonen2013CPC} provides most of the glue code for MPI communication and management of computational domain interfaces.

\begin{figure}
	\centering{
        \includegraphics[trim={12cm 8cm 7cm 3cm}, clip=true, width=12cm]{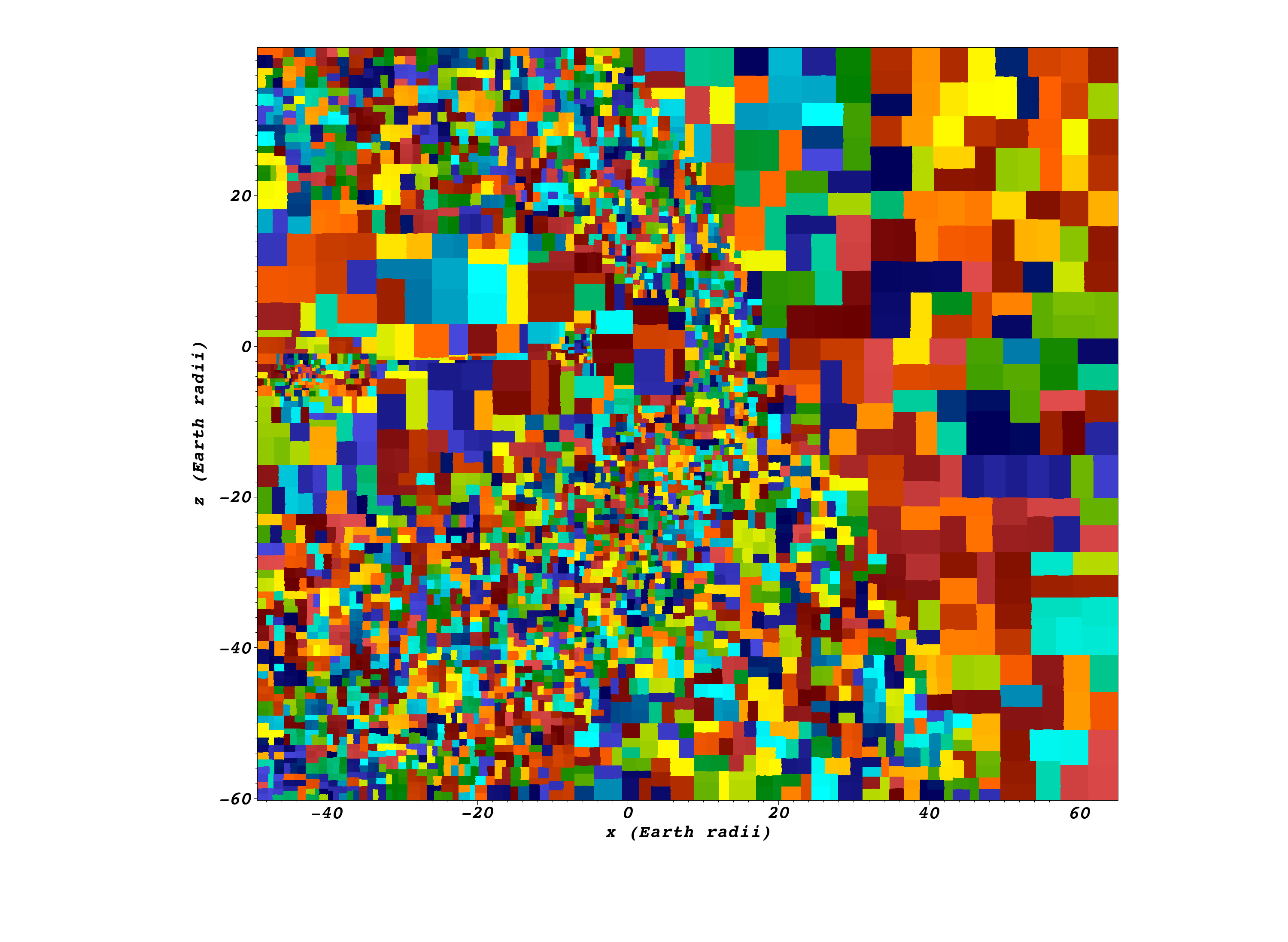}
    }
    \caption{Spatial plot of dynamic load balancing in Vlasiator in a global magnetospheric simulation in the polar plane. The simulation frame is identical to that of Figure \ref{fig:msphere}. Each computational rank 0--4799 is mapped to a colour, and the corresponding rectangular domain is coloured accordingly. Domain decomposition is performed by run-time updated recursive coordinate bisection using the Zoltan library \citep{DevineZoltan2002,BomanZoltan2012}.}
    \label{fig:loadbalancing}
\end{figure}

Thanks to the sparse velocity space representation, large savings in memory usage and computational demand can be achieved. However, the sparse velocity space induces a further problem, because the computational effort to solve the Vlasov equation is no longer constant for every spatial simulation cell, but varies in direct relation to the complexity of the velocity space at every given point. Due to the large variety of physical processes present in the magnetospheric domain, this leads to large load imbalances throughout the simulation box, making a simple Cartesian subdivision of space over computational tasks infeasible and necessitating a dynamic rebalancing of the work distribution. To this end, Vlasiator relies on the Zoltan library \citep{DevineZoltan2002,BomanZoltan2012}, which creates an optimised space decomposition from continuously updated run-time metrics, providing a number of different algorithms to do so (usual production runs are using recursive coordinate bisection). Figure \ref{fig:loadbalancing} shows an example of the resulting spatial decomposition in a 2D global magnetospheric simulation run in the Earth's polar plane, in which the incoming solar wind plasma with low-complexity Maxwellian velocity distributions on the right hand side is processed by visibly larger computational domain sizes than the more complex velocity space structures in the magnetosphere.

The second level of parallelisation is carried out within the computational domain of each MPI task. The domain containing the MPI tasks is typically handled by a full supercomputer node (or a fraction of it) with multiple CPU cores and threads, which includes a local parallelisation level based on OpenMP \citep{OpenMP3.1}.
All computationally intensive solver steps have been designed to be run thread-parallel over multiple spatial cells, or in the case of the SLICE-3D position space update (see section \ref{sec:SLICE3D}) multiple parallel evaluations over the velocity space to make optimum use of the available shared memory parallel computing architecture within one node.  As a third level of parallelisation, all data structures involved in computationally expensive solver steps have been designed to benefit from vector processing of modern CPUs. Specifically, the velocity space representation in Vlasiator is based on $4\times4\times4$ cell blocks, which are always processed as a whole. This allows multiple velocity cells to be solved at the same time, using single instruction multiple data techniques \citep{Agner}.

A further complication of parallel Vlasov simulations are the associated input/output requirements. Not only does it require a parallel input/output system that scales to the required number of nodes, but the sparse velocity space structure requires an appropriate file format able to represent the sparsity, without relying on fixed data offsets. For Vlasiator's specific use case, the VLSV library and file format have been developed (\texttt{http://github.com/fmihpc/vlsv}).
Using parallel MPI-IO \citep[MPI,][]{MPI3.1}, it allows high-performance input/output even for simulation restart files which, given the large system size of Earth's magnetosphere, tend to get up to multiple terabytes in size. A plugin for the popular scientific visualisation suite VisIt \citep{Childs2012Visit} is available, as is a python library that allows for quantitative analysis of the output files (\texttt{http://github.com/fmihpc/analysator}).

Along with the industry's trend towards architectures featuring large numbers of cores and/or GPUs as a primary computing element, an early version of Vlasiator was parallelised using the CUDA standard and run on small numbers of GPUs \citep{Sandroos2013}. This avenue was not pursued further because of the lack of suitably large systems, and a number of bottlenecks following from the structure of the Vlasov simulations on the one hand and the characteristics of GPUs on the other hand.

\subsection{Verification and test problems}

As standard verification tests for a hybrid-Vlasov system do not exist, the first verification effort of Vlasiator was presented in \citet{Kempf2013}. A simulation of low-$\beta$ plasma waves (where $\beta$ is the ratio of thermal and magnetic pressures) in a one-dimensional case with various angles of propagation with respect to the magnetic field was used to generate dispersion curves and surfaces. These were then compared to analytical solutions from the linearised plasma wave equations given by the Waves in Homogeneous, Anisotropic Multicomponent Plasmas (WHAMP) code \citep{Ronnmark1982}. Excellent agreement between the results obtained from the two approaches was found in the case of parallel, perpendicular and oblique propagation, the only noticeable difference taking place for high frequencies and wave numbers, likely as a result of too coarse a representation of the Hall term in the Vlasiator simulations at that time.

In the work presented by \citet{vonAlfthan2014}, the study of the ion/ion right-hand resonant beam instability is another effort to verify the hybrid-Vlasov model implemented in Vlasiator against the analytic solution of the dispersion equation for that instability. The obtained instability growth rates were found to behave as predicted by theory in the cool beam regime, although with slightly lower values which can be explained by the finite size of the simulation box used. This paper also discussed comparisons of results from the hybrid-Vlasov approach with those obtained with hybrid-PIC codes to underline that distribution functions are comparable albeit smoother and better-resolved with the former approach.

More recently, \citet{kilian2017plasma} presented a set of validation tests based on kinetic plasma waves, and discussed what their expected behaviour should look like in fully kinetic PIC simulations as well as different levels of simplification (Darwin approximation, EMHD, hybrid). By nature, waves and instabilities are a sensitive and valuable verification tool for plasma models, as they are an emergence of the collective behaviour of plasma. As such they are an excellent verification test for a complete model going well beyond unit tests of single solver components.

The increasing computational performance of Vlasiator has allowed significant improvements in spatial resolution. It was still 850\,km early on \citep{vonAlfthan2014,Kempf2015} but subsequent runs were performed at 300\,km and even 227\,km resolution \citep[e.g.][]{Palmroth2015,Hoilijoki2017}. Nevertheless even at these finer resolutions the typical kinetic scales are still not properly resolved in magnetospheric simulations. This can lead to the \textit{a priori} concern that under-resolved hybrid-Vlasov simulations would not fare better than their considerably cheaper MHD forerunners and would similarly lack any kinetic plasma phenomena. A systematic study of the effects of the discretisation parameters of Vlasiator on the modelling of a collisionless shock alleviates this concern. Using a one-dimensional shock setup with conditions comparable to the terrestrial bow shock, \citet{PfauKempf2018} show that even at spatial resolutions of 1000\,km the results clearly depart from fluid theory and are consistent with a kinetic description. Of course an increased resolution of 200\,km leads to a dramatic improvement in the physical detail accessible to the model, even though not yet fully resolving ion kinetic scales. This study also highlights the importance of choosing the velocity resolution and the phase-space density threshold $f_\mathrm{min}$ carefully as they affect the conservation properties of the model and as a consequence the physical processes it can describe.

\subsection{Physics results}
\label{sec:vlasiatorScience}

Having completed verification tests, one can compare simulation results with experimental ground-based or spacecraft data, in other words proceed towards validation of the model. The first step for Vlasiator was to perform a global test-Vlasov simulation in 3D ordinary space \citep{Palmroth2013}. In this test $f$ was propagated through the electromagnetic fields computed by the MHD model GUMICS-4 \citep{Janhunen2012}. This test showed that the early test-Vlasov version of Vlasiator already reproduced well the position of the Earth's bow shock as well as magnetopause and magnetosheath plasma properties. Typical energy-latitude ion velocity profiles during northward IMF conditions were also successfully obtained with Vlasiator in that same study.

Focusing on the ion velocity distributions in the foreshock, a study by \citet{Pokhotelov2013} demonstrated that the physics of ions in the vicinity of quasi-parallel MHD shocks is well reproduced by Vlasiator. The simulation presented in that paper is a global dayside magnetospheric run in 2D ordinary space (ecliptic plane) for which the IMF angle relative to the Sun--Earth axis is $45^\circ$. The foreshock was successfully reproduced by the model, and the reflected ion velocity distributions given by Vlasiator were found to be in agreement with spacecraft observations. In particular, deep in the ion foreshock, so-called cap-shaped ion distributions were reproduced by the model in association with 30~s sinusoidal waves which have been created as a result of ion/ion resonance interaction.

Validation of Vlasiator using spacecraft data was presented by \citet{Kempf2015}, where the various shapes of ion distributions in the foreshock were reviewed, localised relative to the foreshock boundaries, identified in Time History of Events and Macroscale Interactions during Substorms \citep[THEMIS,][]{Angelopoulos2008THEMIS} data and compared to model results. The agreement between Vlasiator-simulated distributions and those observed by THEMIS was found to be very good, giving additional credibility to the hybrid-Vlasov approach and its feasibility.

  While the papers discussed above essentially presented validations of the hybrid-Vlasov approach implemented in Vlasiator, the model has since 2015 been producing novel physical results. The first scientific investigations of the solar wind-magnetosphere interaction utilizing Vlasiator focus on dayside processes, from the foreshock to the magnetopause. Vlasiator offers in particular an unprecedented view of the suprathermal ion population in the foreshock. The moments of this population are direct outputs from the code, thus facilitating the analysis of parameters such as the suprathermal ion density or velocity throughout the foreshock \citep{Kempf2015,Palmroth2015}. In contrast, such parameters require some careful data processing to be extracted from spacecraft measurements, and large statistics are needed in order to obtain global maps of the foreshock.

Vlasiator allows to investigate the properties and the structuring of the ultra-low frequency (ULF, 1~mHz to $\sim$1~Hz) waves which pervade the foreshock both on the local and the global scales. Direct comparison of a Vlasiator run with measurements from the THEMIS mission during similar solar wind conditions confirmed that Vlasiator reproduces well the characteristics of the waves at the spacecraft location \citep{Palmroth2015}. The typical features of the waves are in agreement with the reported literature. The observed oblique propagation, relative to the ambient magnetic field, of these foreshock waves has been a long-standing question because theory predicts that they should be parallel-propagating. Based on Vlasiator results, \citet{Palmroth2015} proposed a new scenario to explain this phenomenon, which they attributed to the global variation of the suprathermal ion population properties across the foreshock.

	Vlasiator also offers unprecedented insight into the physics of the magnetosheath, which hosts mirror mode waves downstream of the quasi-perpendicular shock. \citet{Hoilijoki2016} found that the growth rate of the mirror mode waves was smaller than theoretical expectations, but in good agreement with spacecraft observations. As \citet{Hoilijoki2016} explain, this discrepancy has been ascribed to the local and global variations of the plasma parameters, as well as the influence of other wave modes, not being taken into account in the previous theoretical estimates. Using Vlasiator's capability to track the evolution of the plasma as it propagates from the bow shock into the magnetosheath, \citet{Hoilijoki2016} demonstrated that mirror modes develop preferentially along magnetosheath streamlines whose origin at the bow shock lies in the vicinity of the foreshock ULF wave boundary. This is probably due to the fact that the plasma is more unstable to mirror modes in this region because of the perturbations in the foreshock transmitting into the magnetosheath. This result outlines the importance of the global approach, as a similar result would not be present in coupled codes nor in codes that do not model both the foreshock and the magnetosheath simultaneously.

\begin{figure}
	\centering{
		\includegraphics[width=8cm]{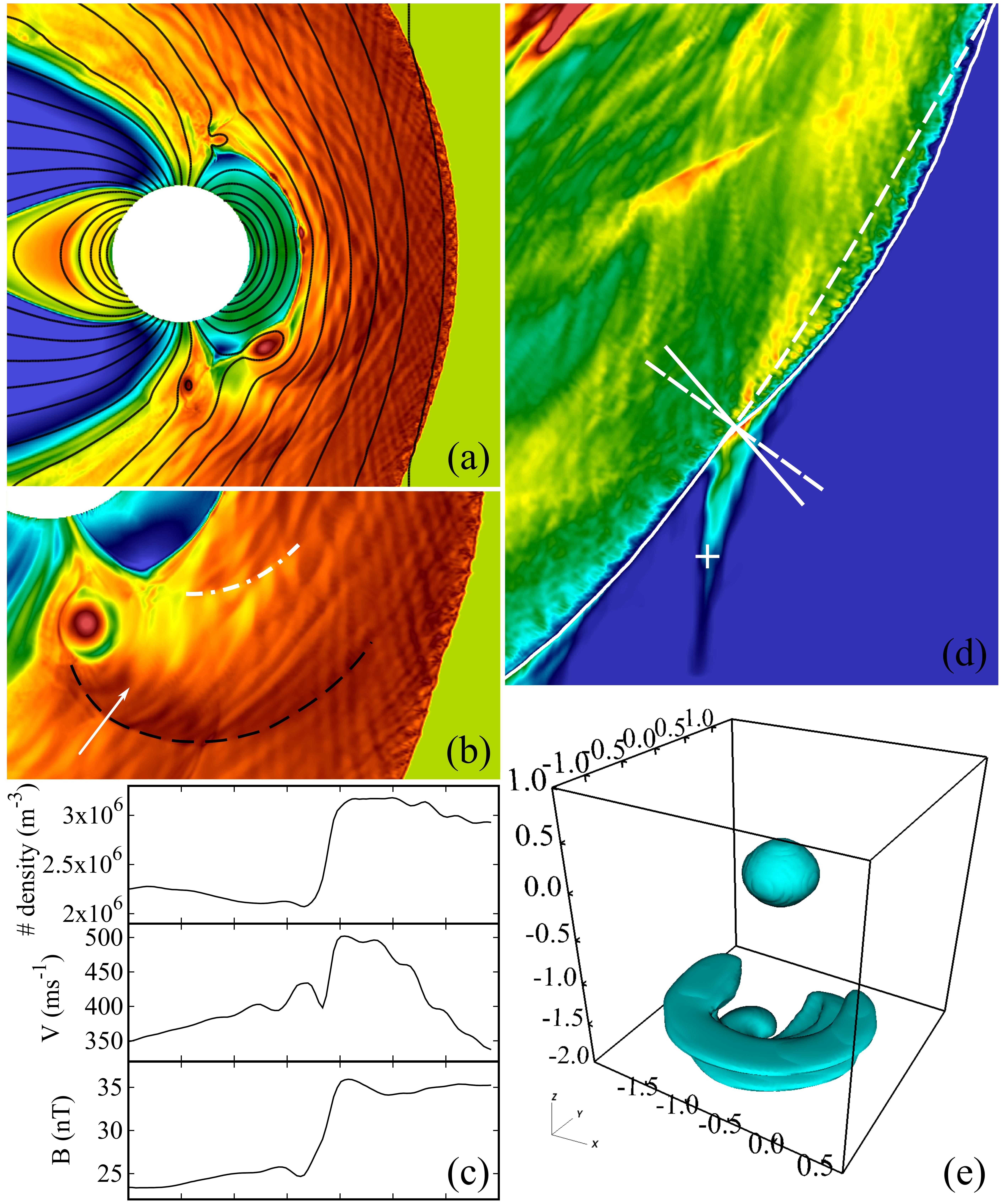}
    }
    \caption{Local, transient foreshocks generated by magnetosheath bow waves. (a) Ion density (colour) and magnetic field lines (black lines) in the near-Earth space, showcasing magnetic islands created by magnetic reconnection. (b) Zoom-in on a strong magnetic island, pushing a bow (dashed black line) and a heck (dashed-dotted white line) wave into the magnetosheath. (c) Plasma parameters across the bow wave, cut along the white arrow in panel b. (d) Parallel ion temperature, showing the transient foreshock as a region of enhanced parallel temperature extending into the solar wind. The bow shock position and its normal direction are indicated by solid white lines, and the dashed lines illustrate the nominal bow shock shape and its normal direction without perturbation. (e) Ion velocity distribution function in the transient foreshock at the location marked by the plus sign in panel d, comparable to what is observed in the regular foreshock. Figure from \citet{PfauKempf2016PhD}.}
    \label{fig:local_foreshock}
\end{figure}

	Magnetic reconnection is a topic of intensive research, as it is the main process through which plasma and energy are transferred from the solar wind into the magnetosphere. Many questions remain unresolved in the dayside and in the nightside. In the dayside, active research topics include the position of the reconnection line and the bursty or continuous nature of reconnection, while in the nightside the most important topic is the global magnetospheric reconfiguration caused either by reconnection or by a tail current disruption. In order to tackle these questions, the simulation domain of Vlasiator, which so far corresponded to the Earth's equatorial plane, was changed to cover the noon-midnight meridian plane ($x-z$ plane in the reference frame defined in section \ref{sec:boundary_conditions}). To address the dayside-nightside coupling processes in reconnection, the simulation domain was extended to include the nightside reconnection site within the same simulation domain, stretching as far as $-94 \, \mathrm{R}_{\mathrm{E}}$ along the $x$ direction. This run, carried out in 2016, remains at the time of this writing the most computationally expensive Vlasiator run performed so far. 
    
    \citet{Hoilijoki2017} presented an investigation of reconnection and flux transfer event (FTE) processes at the dayside magnetopause, and showed that even under steady IMF conditions the location of the reconnection line varies with time, even allowing multiple reconnection lines to exist at a given time. Many FTEs are produced during the simulation, and occasionally magnetic islands have been observed to coalesce, which underlines the power of kinetic-based modelling in capturing highly dynamical and localised processes. Additionally, \citet{Hoilijoki2017} showed that the local reconnection rate measured at locations of the reconnection lines correlates well with the analytical rate for the asymmetric reconnection derived by  \citet{Cassak2007}. This paves the way in using Vlasiator to investigate e.g., the effects of dayside reconnection in the nightside.

	Vlasiator has proven to be a useful and powerful tool to reveal localised phenomena which were never imagined before, and to narrow down the regions of the near-Earth environment where to search for them in observational data sets. One example of this can be found in the work by \citet{PfauKempf2016}, in which transient, local ion foreshocks were discovered at the bow shock under steady solar wind and IMF conditions, as illustrated in Fig. \ref{fig:local_foreshock}. These transient foreshocks were found to be related to FTEs at the dayside magnetopause produced by unsteady reconnection and creating fast mode waves propagating upstream in the magnetosheath (Fig. \ref{fig:local_foreshock}a-c). These wave fronts can locally alter the shape of the bow shock, thus creating favourable conditions for local foreshocks to appear (Fig. \ref{fig:local_foreshock}d and e). Observational evidence giving credit to this scenario was found in a data set comprising Geotail observations near the bow shock and ground-based signatures of FTEs in SuperDARN radar and magnetometer data.

\begin{figure}
	\centering{
		\includegraphics[width=6cm]{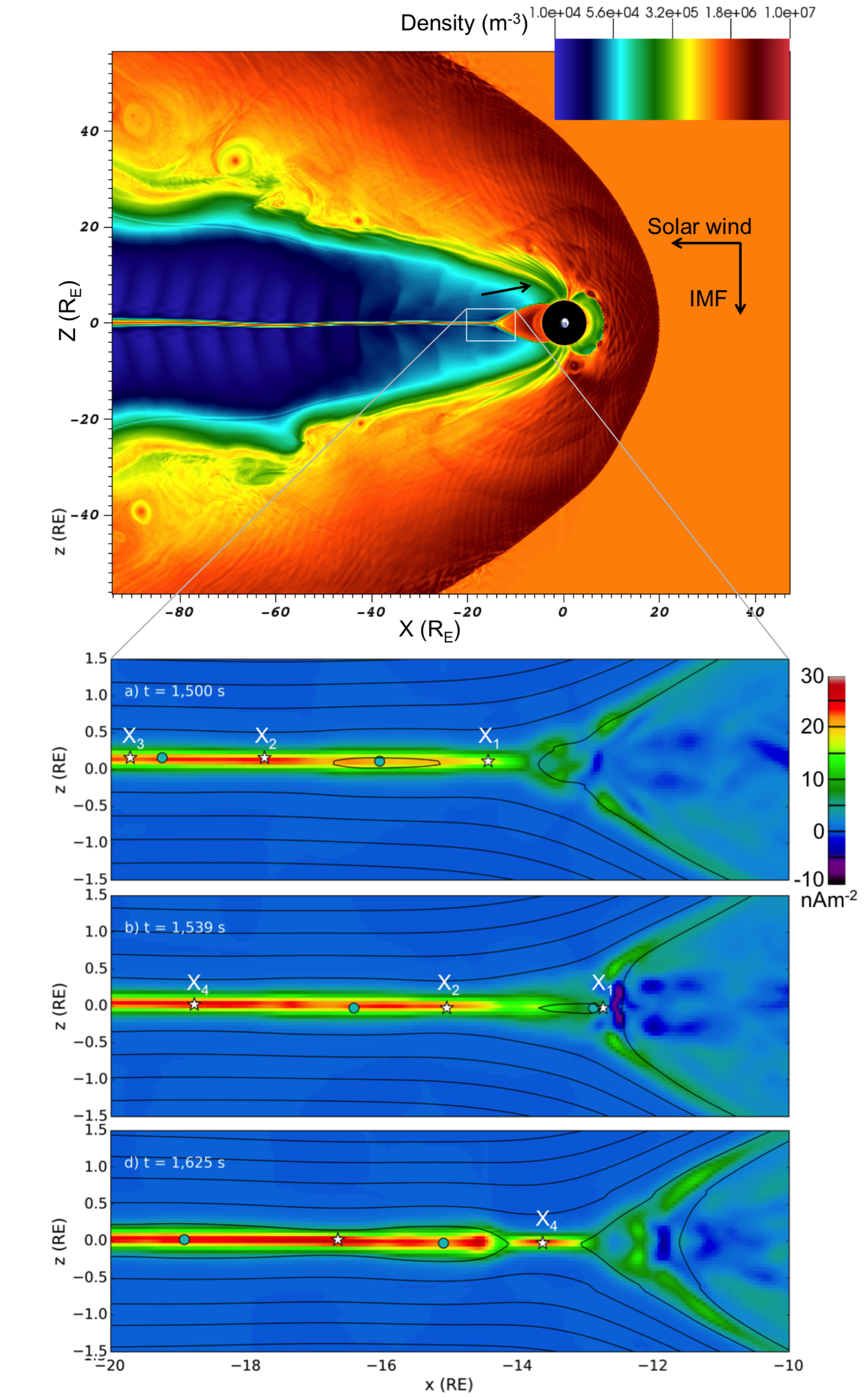}
    }
    \caption{Tail reconnection in Vlasiator. (a) Overview of the ion density in the simulation. (b-d) Close-up of the tail reconnection region at different times in the run.  The colour scheme shows the current density. The X points are marked with stars, which move, appear and disappear as time proceeds. Figure from \citet{Palmroth2017AnGeo}.}
    \label{fig:tail_reconnection}
\end{figure}

	While the first set of publications essentially dealt with dayside processes, Vlasiator can also be applied to the study of nightside phenomena. The first investigation of magnetotail processes using Vlasiator was performed by \citet{Palmroth2017AnGeo}, showcasing multiple reconnection lines in the plasma sheet and the ejection of a plasmoid, under steady IMF conditions (see Fig. \ref{fig:tail_reconnection}). This study underlined that dayside reconnection may have a direct consequence in stabilising nightside reconnection, as flux tubes originating from dayside reconnection influenced the local conditions within the nightside plasma sheet. Again, this study illustrates how important it is to capture the whole system simultaneously using a kinetic approach.

\subsection{Future avenues}
\label{txt:future}
Vlasiator is funded through several multi-annual grants, with which the code is improved and developed. Major building blocks for making Vlasiator possible in the past were not only the increase of the computational resources, but also several algorithmic innovations. Examples of these are the sparse grid for distribution functions, and the semi-Lagrangian solver discussed above. Further, the code has been continuously optimised to fit better in different parallel architectures. With these main steps, the efficiency of Vlasiator has been improved effectively about eight orders of magnitude relative to the performance in the beginning of the project, allowing to simulate 2D-3V systems with high resolution \citep{Palmroth2017AnGeo}. Recently, a simulation run with a cylindrical ionosphere and a layer of grid cells in the third spatial dimension has been carried out, thus approaching the full 3D-3V representation.

The development of Vlasiator is closely tied to the awarded grants. In terms of numerics, near-term plans are to include an adaptive mesh refinement both into the ordinary space and velocity space, required for a full 3D-3V system. These improvements would allow to place higher resolution to regions of interest, and consequently to save in number of time steps and storage. The DCCRG grid already supports adaptive mesh refinement, and thus the task is mainly to add this support into the solvers and optimise the performance. 

In terms of physics, perhaps the most visible change in the near past was the addition of heavier ions. In recent times, the role of heavier ions, e.g., in dayside reconnection has become evident \citep[e.g., ][]{fuselier17}, and thus the correct reproduction of the system at ion scales requires solving heavier ions as well. While the addition requires more memory and storing capacity, in terms of coding it was relatively simple as the ions can be represented as an additional sparse representation of the velocity space, not adding much to the overall computational load. The first runs with heavier ions with additional ion populations were produced in 2018. The first set of runs considered helium flowing from the solar wind boundary, and the second set added oxygen flow from the ionospheric boundary. The analysis for these runs is ongoing.

In the near term, the ionospheric boundary will also be improved. In the 2D-3V runs the ionosphere can be relatively simple, but in 3D-3V it needs to be updated as well. In the first approximation, it can be similar to the type of boundary used in global MHD simulations, which typically couple the field-aligned currents, electric potential and precipitation between the ionosphere and magnetosphere \citep[e.g.][]{Janhunen2012}. Later, the ionosphere should be updated to take into account the more detailed information that the Vlasov-based magnetospheric domain can offer relative to MHD. The objective is to push the inner edge of the simulation domain earthwards from its current position (around 5 $R_E$). Other planned improvements include allowing for the Earth's dipole field to be tilted with respect to the $z$ direction, and replacing the mirror dipole method of ensuring the solenoid condition with an alternative method, for instance a radially vanishing vector potential description of the dipole field. Inclusion of such capabilities would allow investigations of the inner magnetospheric physics in terms of solar wind driving, which would close the circle: The problems of reproducing the inner magnetospheric physics by the global MHD simulations was one of the main motivations for developing Vlasiator in the first place.

Other possible future avenues would be to consider other environments that will be investigated with present and future space missions. An example of this is Mercury targeted by the upcoming BepiColombo mission. Cometary environments and the comet--solar wind interactions  should be interesting in terms of the recently added heavy ion support, in the context of the Rosetta mission data analysis. Further, the upcoming Juice mission will visit the icy moons of Jupiter, indicating that e.g., the Ganymede--Jupiter interaction may also be one viable option for future.

\section{Conclusions and outlook}

There are several main conclusions that can be made from all Vlasiator results so far. The first one is related to the applicability of the hybrid-Vlasov system for ions within the global magnetospheric context. When Vlasiator was first proposed, concerns arose as to whether ions are the dominant species controlling the system dynamics or does one need electrons as well. In particular, a physical representation of the reconnection process may require electrons, while the ion-scale Vlasiator would still model reconnection similarly as global MHD simulations, i.e., through numerical diffusion. However, even an MHD simulation, treating both ions and electrons as a fluid, is capable of modelling global magnetospheric dynamics \citep{palmroth06prec, palmroth06hyst}, indicating that reconnection driving the global dynamics must be within the right ballpark. Since Vlasiator is also able to produce results that are in agreement with \textit{in situ} measurements, kinetic ions seem to be a major contributor in reproducing global dynamics. Whether the electrons play a larger role in global dynamics remains to be determined in the future, if such simulations become possible.

Another major conclusion based on Vlasiator is the role of grid resolution in global setups. Again, one of the largest concerns in the beginning of Vlasiator development was that the ion gyroscales could not be reached within a global simulation volume, raising fears that the outcomes would be MHD-like, even though early hybrid-PIC simulations were also carried out at ion inertial length scales \citep[e.g.,][]{omidi05}. In this context, the first runs included an element of surprise, as even rather coarse resolution grids induce kinetic phenomena that are in agreement with \textit{in situ} observations \citep{Pokhotelov2013}. Latest results have clearly indicated that kinetic physics emerges even at coarse spatial resolution \citep{PfauKempf2018}. It should be emphasised that this result would not have been foreseeable without developing the simulation first. Further, it indicates that also electron physics could be trialled without resolving the actual electron scales. One can hence conclude that others attempting to develop a (hybrid-)Vlasov simulation may face less concerns due to the grid resolution, even in setups with major computational challenges, like e.g., portions of the Sun.

The most common physical conclusion based on Vlasiator simulations is that ``everything affects everything'', indicating that scale coupling is important in global magnetospheric dynamics. One avenue of development for the global MHD simulations in the recent years has been code coupling, where e.g., problem-specific codes have been coupled into the global context \citep{huang06}, or e.g., hybrid-PIC simulations have been embedded within the MHD domain \citep{Toth2016EPICMHD}. While these approaches are interesting and advance physical understanding, they cannot approach scale coupling as the specific kinetic phenomena are only addressed within their respective simulation volumes. A prime example of the scale coupling is the emergence of transient foreshocks, driven by bow waves generated by dayside reconnection \citep{PfauKempf2016}. Another example is the generation of oblique foreshock waves due to a global variability of backstreaming populations \citep{Palmroth2015}. These results could not have been achieved without a simulation that resolves both small and large scales simultaneously.

Vlasov-based methods have not yet been widely adopted in the fields of astrophysics and space physics to model large-scale systems beyond the few examples cited in Table \ref{tab:Applications}, mainly due to the truly astronomical computational cost such simulations can have. The experience with Vlasiator nevertheless demonstrates that Vlasov-based modelling is strongly complementary to other methods and provides unprecedented insight well worth the implementation effort. Based on the pioneering work realised in the Solar-Terrestrial physics community, it is hoped that Vlasov-based methods will gain in popularity and lead to breakthrough results in other fields of space physics and astrophysics as well.

Finally, it should be emphasized that a critical success factor in the Vlasiator development has been the close involvement with technological advances in the field of high-performance computing. European research infrastructures for supercomputing have been developed almost hand-in-hand with Vlasiator, providing an opportunity to always target the newest platforms thus feeding directly into the code development. Should similar computationally intensive codes be designed and implemented elsewhere, it is recommended to keep a keen eye on the technological development of the supercomputing platforms.

\begin{acknowledgements}
We acknowledge The European Research Council for Starting grant 200141-QuESpace, with which Vlasiator (\texttt{http://physics.helsinki.fi/vlasiator}) was developed, and Consolidator grant 682068-PRESTISSIMO awarded to further develop Vlasiator and use it for scientific investigations. We gratefully also acknowledge the Academy of Finland (grant numbers 138599, 267144, and 309937). The Finnish Centre of Excellence in Research of Sustainable Space, funded through the Academy of Finland with grant number 312351, supports Vlasiator development and science as well. We acknowledge all computational grants we have received: PRACE/Tier-0 2012061111 on Hermit/HLRS, PRACE/Tier-1 on Abel/UiO-NOTUR, PRACE/Tier-0 2014112573 on HazelHel/HLRS, PRACE/Tier-0 2016153521 on Marconi/CINECA, CSC -- IT Center of Science Grand Challenge grants on 2015 and 2016, and the pilot use in summer as well as the special Christmas present pilot use of sisu.csc.fi in 2014. LT is supported by Marie Sklodowska-Curie grant agreement No 704681.
\end{acknowledgements}

\bibliographystyle{spbasic}      
\bibliography{bibliography.bib}   

\end{document}